\documentclass{aa}
\usepackage{graphicx}
\usepackage{txfonts}
\begin{document}
   \title{On the efficiency and reliability of cluster mass estimates
   based on member galaxies}

   \author{A. Biviano\inst{1},  G. Murante\inst{2,3}, S. Borgani\inst{3,1,4},
A. Diaferio\inst{5}, K. Dolag\inst{6}, \and M. Girardi\inst{3,1}
          }

\offprints{A. Biviano, biviano@oats.inaf.it}

\institute{INAF/Osservatorio Astronomico di Trieste, via Tiepolo 11, 
34131, Trieste, Italy
\and INAF/Osservatorio Astronomico di Torino, Strada Osservatorio 20, 
10025 Pino Torinese, Italy 
\and Dipartimento di Astronomia dell'Universit\`a di Trieste, via Tiepolo 11, 
34131 Trieste, Italy 
\and INFN - National Institute for Nuclear Physics, Trieste, Italy 
\and Dipartimento di Fisica Generale "Amedeo Avogadro'', Universit\'a 
degli Studi di Torino, Torino, Italy 
\and Max-Planck-Institut f\"ur Astrophysik, Karl-Schwarzschild Strasse 1, 
Garching bei M\"unchen, Germany}

\date{Received / Accepted}

\abstract{}{We study the efficiency and reliability of cluster mass
estimators that are based on the projected phase-space distribution of
galaxies in a cluster region.}{We analyse a data-set of 62 clusters
extracted from a concordance $\Lambda$CDM cosmological hydrodynamical
simulation. We consider both dark matter (DM) particles and simulated
galaxies as tracers of the clusters gravitational potential. Two
cluster mass estimators are considered: the virial mass estimator,
corrected for the surface-pressure term, and a mass estimator (that we
call $M_{\sigma}$) based entirely on the velocity dispersion estimate
of the cluster. In order to simulate observations,
galaxies (or DM
particles) are first selected in cylinders of given radius (from 0.5
to 1.5 $h^{-1}$ Mpc) and $\simeq 200 \, h^{-1}$ Mpc length. Cluster
members are then identified by applying a suitable interloper removal
algorithm.}  {The virial mass estimator overestimates the true mass
by $\simeq 10$\% on average, for sample sizes of $\ga 60$ cluster
members.  For similar sample sizes, $M_{\sigma}$ underestimates the
true mass by $\simeq 15$\%, on average. For smaller sample sizes, the
bias of the virial mass estimator substantially increases, while the
$M_{\sigma}$ estimator becomes essentially unbiased. The dispersion of
both mass estimates increases by a factor $\sim 2$ as the number of
cluster members decreases from $\sim 400$ to $\sim 20$.

It is possible to reduce the bias in the virial mass estimates either
by removing clusters with significant evidence for subclustering or by
selecting early-type galaxies, which substantially reduces the
interloper contamination. Early-type galaxies cannot however be used
to improve the $M_{\sigma}$ estimates since their intrinsic velocity
distribution is slightly biased relatively to that of the DM particles.

Radially-dependent incompleteness can drastically affect the virial
mass estimates, but leaves the $M_{\sigma}$ estimates almost
unaffected. Other observational effects, like centering and velocity
errors, and different observational apertures, have little effect on
the mass estimates.}{}


\titlerunning{On the efficiency and reliability of cluster mass estimates} 
\authorrunning{Biviano et al.}

\maketitle

\section{Introduction}\label{s-intro}
The masses of galaxy clusters are a very useful observable in
cosmology. The number density of clusters of galaxies above a given
mass threshold, and the evolution of this abundance with redshift $z$,
provides in principle strong constraints on cosmological
models/parameters (see, e.g., Rosati et al. \cite{rosa02}; Voit 
\cite{voit05}, for recent reviews, and references
therein). The advantage of using galaxy clusters is that
they are massive, luminous objects, that can be detected relatively
easily out to $z \ga 1$ by several techniques in different
wavebands. The disadvantage is that they are rather complex objects,
hence their masses are not easily estimated, and can often be plagued
by systematic effects uneasy to correct for.

Since Zwicky's (\cite{zwic33}) first estimate of a cluster mass, based
on the application of the virial theorem to the projected phase-space
distribution of galaxies in the Coma cluster, cluster mass
determinations have always been taken with some caution (see the
historical review of Biviano \cite{bivi00}). A cluster mass estimate
based on the observed projected phase-space distribution of galaxies
can be wrong because of several effects. Quite important, in this
respect, are the projection effects leading to the inclusion of
interlopers in the sample of presumed cluster members (see, e.g.,
Lucey \cite{luce83}; Borgani et al. \cite{borg97}; Cen \cite{cen97},
C97 hereafter). Biases in the cluster mass estimates can also occur
when the studied cluster is far from virialization, e.g. during the
accretion phase of a massive group (see, e.g. Girardi \& Biviano
\cite{gira02}), or if galaxies are biased tracers of the gravitational
potential, which could happen as a consequence of dynamical friction
(e.g. Biviano et al. \cite{bivi92}; Goto \cite{goto05}), or as a
consequence of infalling motions (e.g., Moss \& Dickens \cite{moss77};
Biviano et al. \cite{bivi97}).

Because of the above mentioned problems, other methods of cluster mass
determination have been considered, that are not based on the
phase-space distribution of galaxies. While lensing mass estimates are
also known to be affected by projection effects, these are generally
believed to be less severe (e.g. Reblinsky \& Bartelmann
\cite{rebl99}, RB99 hereafter; Clowe et al. \cite{clow04}; but see
Metzler et al. \cite{metz99}), unless the cluster is elongated or
has substantial substructure along the line-of-sight (Athreya et
al. \cite{athr02}; Bartelmann \& Steinmetz \cite{bart96}; Gavazzi
\cite{gava05}; Oguri et al. \cite{ogu05}). Projection effects
are even less important in the case of X-ray emission-based cluster
mass estimates, since the X-ray emissivity is proportional to the
square of the gas density.

Problems with lensing and X-ray mass estimates do exist,
however. Masses determined with the lensing technique are affected by
the mass-sheet degeneracy, that cannot always be broken (e.g. Dye et
al. \cite{dye01}; Cypriano et al. \cite{cypr04}). The effect of
intervening matter along the line-of-sight is to increase the
weak-lensing mass estimates of clusters, especially at high redshifts
(Metzler et al. \cite{metz01}; Lombardi et al. \cite{lomb05};
Wambsganss et al. \cite{wamb04,wamb05}).  Lensing and X-ray-based
mass estimates have often been found to be discordant, and this
is generally interpreted as evidence of non-equilibrium (e.g. Wu
\cite{wu00}; Clowe et al. \cite{clow01}; Athreya et al. \cite{athr02};
Smith et al. \cite{smit02}; Cypriano et al. \cite{cypr04}; Ota et
al. \cite{ota04}; Brada\v{c} et al. \cite{brad05}). In clusters
undergoing merging events the X-ray luminosity and temperature can be
boosted, thus leading to an overestimate of the cluster mass
(e.g. Schindler \& M\"uller \cite{schi93}; Ricker \& Sarazin
\cite{rick01}; Barrena et al. \cite{barr02}; Diaferio et
al. \cite{diaf05}). On the other hand, violations of the condition of
hydrostatic equilibrium, inaccurate modelling of the gas density
profile and observational biases in the measure of the intra--cluster
gas temperature, may lead to a sizeable underestimate of the cluster
mass (e.g. Bartelmann \& Steinmetz \cite{bart96}; Kay et
al. \cite{kay04}; Rasia et al. \cite{rasi04}).

A combination of several, independent cluster mass estimates is likely
to provide the most accurate results. Moreover, with the Sloan Digital
Sky Survey coming to completion (e.g. Abazajian et al. \cite{abaz05}),
a large number of nearby clusters with $\geq 30$ galaxies with
redshifts is now available (Miller et al. \cite{mill05}). For many of
these clusters, X-ray data are not available, and mass estimates must
be based on optical data. Hence, mass estimates based on the projected
phase-space distribution of galaxies are still very useful.

Time is therefore mature for a re-assessment of the reliability of
mass estimates of clusters based on the dynamics of their member
galaxies. Previous analyses have generally considered only specific
aspects of this topic. Initially, the reliability of different mass
estimators has been assessed from N-body simulations outside of a
cosmological context (Danese et al. \cite{dane81}; Perea et
al. \cite{pere90}; Aceves \& Perea \cite{acev99}). In order to
properly deal with this topic, clusters must however be identified
within cosmological simulations with large box-sizes. This has been
achieved by several studies (e.g. Frenk et al. \cite{fren90}; Borgani
et al. \cite{borg97}; C97; van Haarlem et al. \cite{vanh97}, hereafter
vH97; RB99; Sanchis et al. \cite{sanc04}; \L okas et
al. \cite{loka05}), where, however, dark matter (DM hereafter)
particles and not galaxies, were considered as tracers of the
potential. Van Kampen \& Katgert (\cite{vank97}) used the method of
constrained random fields to increase the numerical resolution of
their simulation, but by doing this they were unable to consider
projection effects.  In most analyses, interlopers were rejected using
Yahil \& Vidal's (\cite{yahi77}) traditional 3-$\sigma$-clipping
method (but see, e.g., vH97 who also tested the more sophisticated
method of den Hartog \& Katgert \cite{denh96}, and Sanchis et al.
\cite{sanc04} and \L okas et al. \cite{loka05} who tested their own
methods of interloper removal).

A general conclusion from these studies is that cluster mass estimates
can be severely affected by projection effects. This happens mainly as
a consequence of the cluster identification process in 2-d projected
space, performed with Abell's (\cite{abel58}) original
algorithm. Frenk et al. (\cite{fren90}) argued that cluster masses are
systematically over-estimated, but following studies (C97; vH97; RB99)
concluded that cluster masses can be either over- or under-estimated,
depending on the projection angle, the cluster mass, and the algorithm
used to remove interlopers. Sanchis et al. (\cite{sanc04}) and \L okas
et al. (\cite{loka05}) found a rather good agreement between estimated
and true cluster masses, when a rather large number of tracers of the
potential was considered in each cluster.

None of the above mentioned studies tried to identify galaxies in the
cosmological simulations. In those studies where this was achieved, it
was generally found that the spatial distribution of {\em subhaloes,}
selected by their mass, is less concentrated than that of DM
(e.g. Ghigna et al. \cite{ghig98}; Klypin et al. \cite{klyp99}; Gao et
al. \cite{gao04}). However, when gasdynamics was included in the
simulations, {\em galaxies,} selected by luminosity, turned out to have
a considerably more concentrated spatial distribution than subhaloes,
and more similar to that of DM (Berlind et al. \cite{berl03}; Gao et
al. \cite{gao04}; Nagai \& Kravtsov \cite{naga05}). This occurs
because tidal stripping induce substantial mass loss from galaxy
haloes, but very little stellar mass loss.

As far as the velocity distribution of {\em subhaloes} is concerned,
most studies have found it to be wider than that of DM particles
(Col\'{\i}n et al. \cite{coli00}; Diemand et al. \cite{diem04}), at
least near the cluster centre (Ghigna et al. \cite{ghig00}; Reed et
al. \cite{reed05}). Based on hydrodynamical simulations, Frenk et
al. (\cite{fren96}) however concluded that {\em galaxies} suffer
significant dynamical friction, and are slowed down relatively to DM
particles, to such an extent that cluster mass estimates based on the
velocity dispersion of galaxies are likely to be in error by factors
of 0.25--0.75, depending on the masses of the galaxies selected as
tracers. On the other hand, Berlind et al. (\cite{berl03}) found that
galaxies have only a mild velocity bias with respect to DM, and
Faltenbacher et al. (\cite{falt05}) found that galaxies, if anything,
move slightly faster than DM particles.

In this paper the reliability of cluster mass estimates based on the
dynamics of their member galaxies is reconsidered on the basis of a
set of clusters extracted from a large cosmological hydrodynamical
simulation (Borgani et al. \cite{borg04}; see also Murante et
al. \cite{mura04}), performed using the TREE+SPH {\small GADGET--2}
code (Springel et al. \cite{spri01}; Springel \cite{spri05}), for a
concordance $\Lambda$CDM model.  This simulation samples a fairly
large volume (box size of 192 $h^{-1}$ Mpc, where $h$ is the Hubble
constant in units of 100~km~s$^{-1}$~Mpc$^{-1}$) which contains more
than 100 clusters with mass above $10^{14}h^{-1} M_{\odot}$, while
also having good enough resolution ($m_{\rm DM} = 4.62 \times 10^9
h^{-1}M_\odot$ for the mass of the DM particles) to allow resolving
halos hosting bright galaxies. Finally, the inclusion of the processes
of radiative cooling, star formation and supernova feedback allows us
to have a realistic description of the gas evolution and of the galaxy
formation process.

In this paper we address the question of how accurate are cluster mass
determinations based on the dynamics of their member galaxies, under a
variety of observational conditions, and for a wide range of cluster
masses, but independently of the clusters identification procedure. No
attempt is made here to simulate the observational identification
algorithms of galaxy clusters, such as Abell's (\cite{abel58})
original one. In this sense, our approach is different from previous
ones (Frenk et al. \cite{fren90}; C97; vH97; RB99), in that we
disentangle the problem of clusters mass estimation from that of
clusters identification. The rationale for this choice is that today
there is {\em no} standard cluster identification algorithm. Automated
scans of digitized plates have since replaced Abell's eyeball
identification of galaxy clusters (Dalton et al. \cite{dalt92};
Lumsden et al. \cite{lums92}; Lopes et al. \cite{lope04}), and much
more sophisticated algorithms than Abell's (\cite{abel58}) have been
applied for extracting the 2-dimensional signal produced by a galaxy
overdensity (e.g. Ramella et al. \cite{rame01}; Gladders \& Yee
\cite{glad05}, and references therein). Moreover, those clusters whose
masses are derived using member galaxies, are not necessarily
optically selected (e.g. Popesso et al. \cite{pope05}). The results of
this paper are therefore useful for a better understanding of why
different techniques (lensing, X-ray, galaxies) can lead to discrepant
cluster mass estimates, and also for the study of scaling relations of
different cluster properties with cluster masses. Translating the
results of this paper to the study of the distribution of observed
cluster masses in a given survey, requires convolution with the
selection function of the survey itself.

The plan of this paper is as follows.  In Sect.~\ref{s-sims} the set
of simulated clusters and its characteristics are described. In
Sect.~\ref{s-mass} the steps involved in the determination of cluster
dynamical quantities are briefly described. The analyses of the
cluster dynamics in projected phase-space are described in
Sect.~\ref{s-2d}. Results are discussed and summarized in
Sect.~\ref{s-disc}. In Sect.~\ref{s-summ} we provide our conclusions.

\section{The simulated clusters}\label{s-sims}
The set of simulated clusters, analysed in this paper, are extracted
from a large cosmological hydrodynamical simulation presented by
Borgani et al. (\cite{borg04}), and performed using the massively
parallel Tree+SPH {\small GADGET--2} code (Springel et
al. \cite{spri01}; Springel \cite{spri05}). We refer to Borgani et
al. (\cite{borg04}) for a complete description of this simulation and
of the cluster identification procedure. We provide here their main
characteristics and describe in some details the galaxy identification
procedure. The simulation assumes a cosmological model with
$\Omega_0=0.3$, $\Omega_\Lambda=0.7$, $\Omega_{\rm
b}=0.019\,h^{-2}$, $h=0.7$, and $\sigma_8=0.8$. The box size is $192
h^{-1}$ Mpc. We used $480^3$ DM particles and (initially) as many gas
particles, for a mass resolution of $m_{\rm DM}=4.62 \times 10^9\,
h^{-1} M_\odot$.  The Plummer--equivalent softening length was set to
$\epsilon=7.5\, h^{-1}$ kpc at $z=0$. The simulation code includes
explicit energy end entropy conservation, radiative cooling, a uniform
time-dependent UV background (Haardt \& Madau \cite{haar96}), the
self-regulated hybrid multi-phase model for star formation (Springel
\& Hernquist \cite{spri03}), and a phenomenological model for galactic
winds powered by Type-II supernovae.

We identify galaxy clusters at redshift $z=0$ by applying a standard
Friends-of-friends (FoF) analysis to the DM particle set, with linking
length $\lambda=0.15$ in units of the mean interparticle
distance. After the FoF identification, we applied a spherical
overdensity criterion to find the virial region of each cluster
(corresponding to an overdensity of $\approx 100$ times the critical
density for the adopted cosmology). We identify 117 galaxy clusters
whose virial mass is larger than $10^{14}\,h^{-1} M_\odot $ within our
simulated volume. A subset of these clusters is selected for the
analysis discussed in the following. The position of the minimum
potential particle of each FoF group is then used as the centre for a
spherical overdensity algorithm, which identifies the radii
encompassing different overdensities.

Galaxies are identified using the publicly available
algorithm\footnote{See \\ {\tt
http://www-hpcc.astro.washington.edu/tools/skid.html}} SKID (Stadel
\cite{stad01}). SKID calculates a density field using all the
particles. Density is assigned with a spline-kernel similar to that
employed by SPH codes; an important parameter is the number of
neighbouring particles $N_{sm}$ on which the density is softened. 
Particles are then associated to the local maxima of such field, which
should represent the positions of the substructure. To this aim,
particles are ``moved'' along the gradient of the field until they
begin to oscillate around the peaks and are then grouped using a FoF
analysis on the ``moved'' positions. An unbinding procedure is 
finally applied on the resulting FoF groups, to discard particles
which are not gravitationally bound to the group. All particles within
a sphere of given radius are used to evaluate the gravitational
potential. This step is performed on the ``true'' (i.e. not ``moved'')
positions. All the length scales involved in the process are set
starting from the parameter $\tau$ which represents the typical size
of the objects to be found. Apart from the evaluation of the density
field, it is possible to select to which types of particles (DM,
stars, and gas) the algorithm must be applied. We used only star
particles to identify our galaxies.

The problem of reliably detecting sub-structures within given DM
haloes is long-lasting and still not uniquely solved. It is outside
the purpose of the present work to discuss such an issue. We performed
a number of tests on the SKID galaxies, to verify that the least
possible number of objects is missed by the analysis and that the
identified object are real ones. Such tests are presented elsewhere
(Murante et al, in preparation). Our galaxies have been identified by
using $\tau = 20 h^{-1}$ kpc, which, after a trial-and-error
procedure, turned out to be the optimal choice. This choice is also
motivated by the fact that $20 h^{-1}$ kpc roughly correspond to the
effective force resolution of the simulation. We also determined that
distinct SKID analyses are needed, with different values of $N_{sm}$,
that we assumed to be 16, 32 and 64. Some galaxies are ``missed''
using only one value for $N_{sm}$. Therefore, we built our catalogue
by joining together the results of the three SKID analyses, with the
following rule: if a star particle is a member of a galaxy for one
value of $N_{sm}$, it is considered to belong to that galaxy; if a
particle is member of two different galaxies for two different values
of $N_{sm}$, the two galaxies are then merged into a single
object. SKID objects with less than 32 particles are discarded.

For the purpose of this paper it is not useful to consider clusters
with less than 10 galaxies, hence in the following we restrict our
analysis to those 62 clusters with at least 10 galaxies within a
sphere of 'virial' radius $r_v$, defined as the radius where the mass
density of the cluster equals 200 times the critical density of the
Universe.

In Borgani et al. (\cite{borg04}) we have shown that these simulated
clusters have on average a star fraction which is $\sim 50$ per cent
larger than the typical observed values (Lin, Mohr \& Stanford
\cite{lin03}). However, this overcooling is mainly contributed by the central
galaxy. Therefore, our simulations are expected to reliably describe
the way in which cluster galaxies trace the underlying cluster
dynamics.

\begin{figure}
\centering \resizebox{\hsize}{!}{\includegraphics{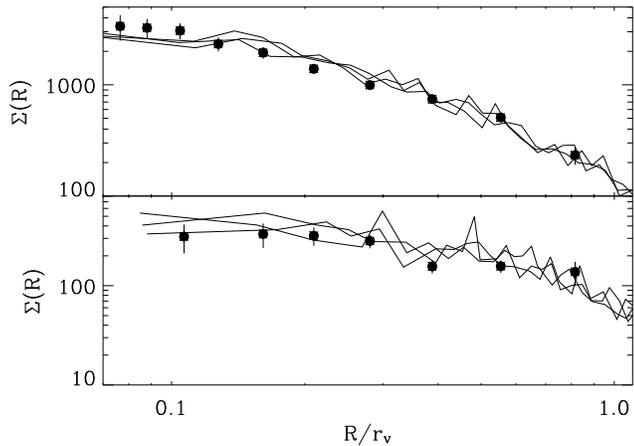}}
\caption{The projected number density profiles of simulated galaxies
(solid lines) and real galaxies from the ENACS sample (dots with error
bars). The three solid lines in each plot correspond to three
orthogonal projections. The normalisations of the simulated galaxies
profiles have been arbitrarily scaled to match the observed
profiles. The clustercentric distances on the x-axis are in units of
the cluster virial radius.  Top panel: simulated galaxies with
formation redshift $z_f \geq 1.25$ (see text) vs. early-type galaxies
from the ENACS sample. Bottom panel: simulated galaxies with $z_f < 
1.25$ vs. late-type galaxies from the ENACS sample.}
\label{f-sigmar}
\end{figure}

\begin{figure}
\centering \resizebox{\hsize}{!}{\includegraphics{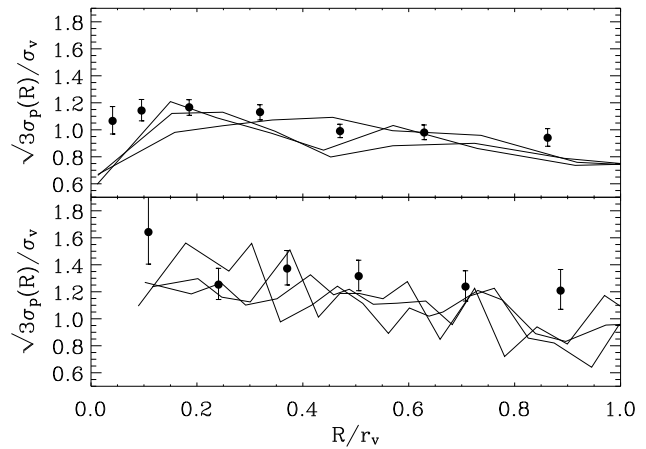}}
\caption{The line-of-sight velocity dispersion profiles (in units of
the total cluster velocity dispersion) of simulated galaxies (solid
lines) and real galaxies from the ENACS sample (dots with error
bars). The clustercentric distances on the x-axis are in units of the
cluster virial radius. The three solid lines in each plot correspond
to three orthogonal projections. Top panel: simulated galaxies with
formation redshift $z_f \geq 1.25$ (see text) vs. early-type galaxies
from the ENACS sample. Bottom panel: simulated galaxies with $z_f <
1.25$ vs. late-type galaxies from the ENACS sample.}
\label{f-vdp}
\end{figure}

For each galaxy, we define its formation redshift, $z_f$, as the
average value of the formation redshifts of all its member star
particles. The value of $z_f$ is also used to classify our galaxies
into ``early--type'' and ``late--type''. Specifically, we consider a
galaxy to be of early-type if $z_f \geq 1.25$, and of late-type if
$z_f<1.25$. This choice is admittedly rather crude and is based on the
comparison of our simulated galaxy sample with the sample of cluster
galaxies of the ESO Nearby Abell Clusters Survey (ENACS, Katgert et
al. \cite{katg96}, \cite{katg98}). In the ENACS, 64\% of the galaxies
identified as cluster members within $1.5 \, r_v$ are classified as
early-type galaxies (ellipticals, lenticulars, or an intermediate
class between these two; see Biviano et al. \cite{bivi02}; Thomas \&
Katgert \cite{thom06}). Similarly, in the simulated clusters, 64\% of
the galaxies within $1.5 \, r_v$ have $z_f \geq 1.25$.

Further support to the identification of $z_f \geq 1.25$ galaxies as
early-type galaxies comes from the comparison of their distribution in
projected phase-space with that observed for the ENACS cluster
galaxies. Since the number of galaxies per simulated cluster is rather
limited, clusters are stacked together, by scaling the galaxy
clustercentric distances and velocities by the value of $r_v$ and the
velocity dispersion of the cluster they belong to. A similar procedure
is applied to the observed galaxies (see Katgert et al. \cite{katg04},
and Biviano \& Katgert \cite{bivi04}).  The projected number density
profiles, $\Sigma(R)$, and the line-of-sight velocity dispersion
profiles, $\sigma_p(R)$, of real and simulated galaxies are shown in
Figures~\ref{f-sigmar} and \ref{f-vdp}, respectively, separately for
early- and late-type galaxies (top and bottom panels, respectively).
The simulated galaxies $\Sigma(R)$'s have been arbitrarily rescaled to
match the observed $\Sigma(R)$'s, since here we are only interested in
the {\em relative} distributions of simulated and real galaxies.

The profiles of observed and simulated galaxies are rather similar,
both for the early- and the late-types. The main difference
between the observed and simulated profiles is the lower number
density and smaller velocity dispersion of the simulated early-type
galaxies in the inner regions ($R/r_v \la 0.1$). This difference is
caused to the well-known overmerging problem which affects the
simulations in the denser regions. While this can be a source of
concern when comparing simulated vs. observed cluster mass {\em
profiles,} we argue that the analysis of cluster masses is not
significantly influenced by this problem. In fact, the apparent
underdensity of simulated early-type galaxies in the inner regions, as
compared to the real ones, is not really significant. The fraction of
early-type galaxies within the virial region, that lie within $0.1 \,
r_v$ is 12.9\% for the simulated galaxies, and 14.5\% for the observed
ones. As a consequence, no significant effect is expected on cluster
harmonic mean radius estimates (which enters the virial mass estimate,
see Sect.~\ref{s-mass}), also because these fractions are low.  For
the same reason, no significant effect is expected on cluster global
velocity dispersion estimates, despite the fact that the difference in
the velocity dispersions of simulated and observed early-type galaxies
{\em is} significant within $0.1 \, r_v$. In fact, the ratio between
the velocity dispersion estimates obtained using all galaxies and only
those with $R > 0.1 \, r_v$ is 0.994 for the simulated galaxies, and
1.008 for the real ones.

We conclude that the similarity between the profiles of simulated and
observed galaxies lends support to our choice of $z_f=1.25$ for
separating early- and late-type galaxies in our simulations.

\section{The cluster mass estimates}\label{s-mass}
Several different definitions of 'virial mass' have been given in the
literature. This term may be used to define the total mass of a
cluster within a radius of given overdensity, i.e. the 'virial
radius', typically the radius where the mean cluster mass density
equals 100 or 200 times the Universe critical density (see, e.g., \L
okas \& Mamon \cite{loka03}). On the other hand, the same term is used
to define the mass estimated through application of the virial theorem
to the cluster galaxies within an observationally defined aperture
(e.g. Biviano et al. \cite{bivi93}; Girardi et
al. \cite{gira98}). This estimate requires not only an estimate of the
cluster velocity dispersion, but also of the harmonic mean radius of
the spatial distribution of cluster galaxies. To further complicate
the issue, twice the harmonic mean radius is also usually referred to
as the 'virial radius' (e.g. Girardi et al. \cite{gira98}; Merch\'an
\& Zandivarez \cite{merc05}).

The virial mass estimate requires correction for the surface pressure
term (The \& White \cite{the86}), unless the {\em entire} system is
contained within the observationally defined aperture radius (see also
Macci\`o et al. \cite{macc03}). Unfortunately, it is still relatively
uncommon to see this correction applied (see, e.g., Koranyi et
al. \cite{kora98}). Neglecting this correction leads to overestimate
the mass of a system (see, e.g., Carlberg et al. \cite{carl97a}), and
this can partly account for some of the claimed discrepancies between
optically and X-ray derived cluster mass estimates.

For the sake of clarity, we detail in the following the procedure that
we apply to our simulated clusters in projection in order to simulate
an observational estimate of their masses. Such a procedure has
recently been applied to a large set of nearby clusters by Popesso et
al. (\cite{pope05}, \cite{pope06}).

\begin{figure}
\centering
\resizebox{\hsize}{!}{\includegraphics{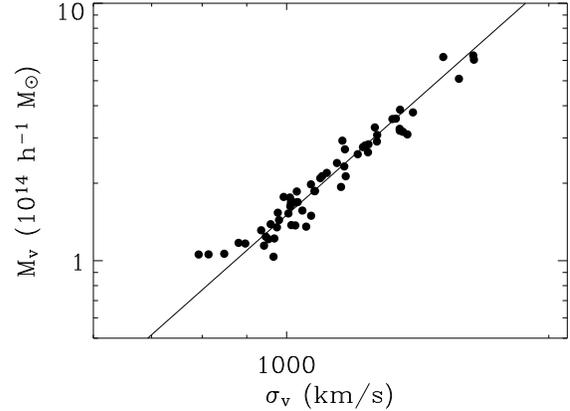}}
\caption{The true mass vs. the 3-d velocity dispersion (both computed
within a sphere of radius $r_v$) for the 62 simulated clusters
with at least 10 galaxies within $r_v$. The best fitting cubic law
relation is shown (solid line).}
\label{f-mvsv}
\end{figure}

\begin{figure}
\centering
\resizebox{\hsize}{!}{\includegraphics{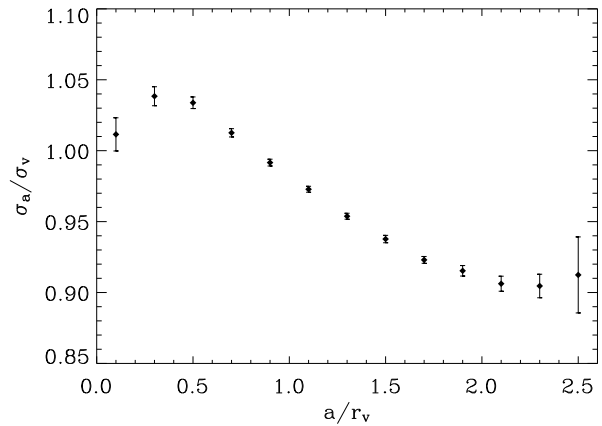}}
\caption{The average integrated velocity dispersion profile for the 62
simulated clusters. The 3-d velocity dispersion within a sphere of radius
$a$ is shown as a function of $a/r_v$, in units of $\sigma_v$. 1-$\sigma$
error bars are shown.}
\label{f-sigpint}
\end{figure}

We define a cluster 'true' mass, $M_v$, as the total mass within the
radius $r_v$, where the mass density of the cluster equals 200 times
the critical density of the Universe at the cluster redshift (we call
$\sigma_v$ the 3-d velocity dispersion of the DM particles within the
same radius). The virial mass estimate $\tilde{M}_v$ within the same
radius is computed as follows.
\begin{enumerate}
\item Define an aperture radius, $a$, within which to perform the
dynamical analysis, and select all DM particles or all galaxies within
a cylinder of radius $a$ and $\simeq 200 \, h^{-1}$ Mpc length, along
each of three orthogonal projections. Cluster centres are computed as
the positions of the minimum potential particle in each cluster (see
Sect.~\ref{s-sims}), unless otherwise indicated (see
Sect.~\ref{s-others}). In virtually all cases, these centres
correspond (to within $\sim 10$ kpc) with the peaks of the cluster
X-ray emissivity. The same centre choice is usually adopted
observationally, when X-ray data are available (see, e.g., Biviano et
al. \cite{bivi97}; Popesso et al. \cite{pope05}).
\item Select cluster members. This is done first by using a cut in
line-of-sight velocity space of $\pm 4000$~km~s$^{-1}$ with respect to
the mean cluster velocity, initially defined by applying the biweight
estimator\footnote{Here and throughout this paper we use the biweight
estimator for the average and dispersion, unless otherwise indicated.}
(see Beers et al. \cite{beer90}) to all the objects within 0.5
$h^{-1}$ Mpc from the cluster centre. In a second step, the weighted
gap procedure of Girardi et al. (\cite{gira93}) is applied on the
remaining objects. On the objects that pass the weighted gap selection
we finally apply Katgert et al.'s (\cite{katg04}) procedure that makes
use of the location of galaxies (or particles) in projected
phase-space (see also den Hartog \& Katgert's \cite{denh96}).
\item Determine the 'robust' estimate of the line-of-sight velocity
dispersion $\sigma_{a,p}$, within the aperture $a$, using the biweight
or gapper estimator, depending on the number of data available, $\geq
15$ or $< 15$, respectively (see Beers et al. \cite{beer90}).
\item Correct the velocity dispersion estimate for velocity errors (if
these are added to the simulated data, see Sect.~\ref{s-others})
following the prescriptions of Danese et al. (\cite{dane80}).
\item Determine the projected harmonic mean radius $r_{a,p}$, within
the aperture $a$.
\item Obtain a first estimate of the mass from
\begin{equation}
M_a \equiv 3 \pi \sigma_{a,p}^2 r_{a,p} / G, 
\end{equation}
where $G$ is the gravitational constant, the factor $3 \pi/2$ is the
deprojection factor, see Limber \& Mathews (\cite{limb60}), and a
factor 2 is needed to convert the harmonic mean radius into the
'virial radius' of Girardi et al. (\cite{gira98}).
\item Estimate the Navarro et al. (\cite{nava97}, NFW hereafter)
concentration parameter using the relation $c=4
(\sigma_{a,p}/700)^{-0.306}$.  The normalization of the relation is
taken from Katgert et al. (\cite{katg04}), and the exponent of the
relation is derived from Dolag et al. (\cite{dola04}), under the
assumption that the mass scales with the third power of the velocity
dispersion.
\item Correct the mass estimate for the surface pressure term,
$f_{sp}$, obtained for a NFW profile with concentration $c$: $M_{a,c}
\equiv f_{sp} M_a$, assuming isotropic orbits (see eq.(8) in
Girardi et al. \cite{gira98}).  Note that the value of the correction
factor is not negligible ($f_{sp}=0.84$ on average for our simulated
clusters observed out to an aperture radius $a=1.5 \, h^{-1}$~Mpc).
\item Determine $\tilde{r}_v$, an estimate of $r_v$, as $\tilde{r}_v=a
[\rho_a/(200 \rho_c(z))]^{1/\xi}$, where $\rho_c(z)$ is the critical
density of the Universe at the cluster redshift, $\rho_a \equiv M_{a,c} / (4
\pi a^3 /3)$, and $\xi$ is the local slope of a NFW profile of
concentration $c$ at the radius $a$.
\item Determine $\tilde{M}_v$, an estimate of $M_v$, by extrapolating or
interpolating $M_{a,c}$ from $a$ to $\tilde{r}_v$ using a NFW profile with
concentration $c$.
\end{enumerate}

\begin{figure}
\centering
\resizebox{\hsize}{!}{\includegraphics{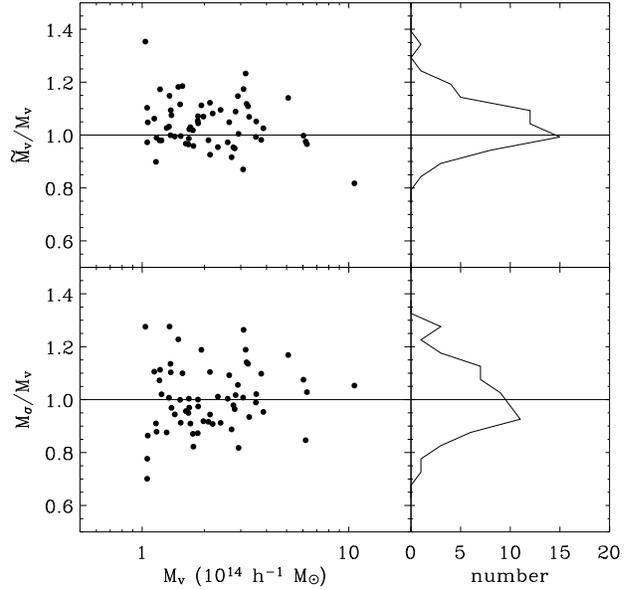}}
\caption{Verifying the performance of the mass estimators: the
comparison between the masses estimated using full phase-space
information, and the true cluster masses.  Top-left panel: ratio
between the virial and true mass, $\tilde{M}_v/M_v$ vs. the true mass,
$M_v$. Bottom-left panel: ratio between the mass obtained from
$\sigma_v$ and the true mass, $M_{\sigma}/M_v$ vs. the true mass
$M_v$.  Top-right panel: distribution of
$\tilde{M}_v/M_v$. Bottom-right panel: distribution of
$M_{\sigma}/M_v$.  The masses $\tilde{M}_v$ and $M_{\sigma}$ are
estimated from the full phase-space distributions of 5000 DM particles
per system (when available; otherwise all DM particles are selected)
within the cluster virial radius.}
\label{f-dm5000}
\end{figure}

Quite often, cosmological constraints have been obtained directly from
the distribution of cluster velocity dispersions, used as proxies for
the cluster masses (see, e.g., Girardi et al. \cite{gira93}). It is
therefore useful to also consider an alternative (and simpler) mass
estimator entirely based on the line-of-sight estimate of the velocity
dispersion within a given aperture, $\sigma_{a,p}$. The relation
between $M_v$ and $\sigma_v$ for our simulated clusters is shown in
Fig.~\ref{f-mvsv}. The line in that plot represents the best-fitting
cubic relation
\begin{equation}
M_v \equiv A \, \left({\sigma_v \over 10^3 {\rm
    \,km\,s^{-1}}}\right)^3\times 10^{14} \, h^{-1} \, M_{\odot} 
\label{m-sig}
\end{equation}
where $A=1.50 \pm 0.02$ (1-$\sigma$ error). We note that this
relation is approximately independent of the cosmological model
(Borgani et al. \cite{borg99}).  It is possible to use this relation
to obtain a mass estimate entirely based on the velocity dispersion
estimate. Since $\sqrt{3} \sigma_{a,p}$ can differ significantly from
$\sigma_v$ depending on the aperture radius $a$ chosen by the
observer, we need to apply a correction that depends on the ratio
$a/r_v$. For each of the 62 clusters, we determine the 3-d velocity
dispersion $\sigma_a$, within a sphere of radius $a$, for different
values of $a$. The average $\sigma_a/\sigma_v$ vs. $a/r_v$ profile of
our 62 clusters is shown in Fig.~\ref{f-sigpint}.

Hence, given a line-of-sight velocity dispersion estimate within a
given observational aperture, $\sigma_{a,p}$, we proceed as follows.
We use $\sqrt{3} \sigma_{a,p}$ to determine an initial estimate of the
mass of the system through eq.~\ref{m-sig}. We then obtain an estimate
$\tilde{r}_v$ of $r_v$, following steps 7 and 9 above (note that step
8 is not needed in this case, because the relation provided by
eq.~\ref{m-sig} relates the velocity dispersion to the {\em true}
cluster mass, for which no surface term correction is needed). Using
the $\sigma_a/\sigma_v$ vs. $a/r_v$ profile of Fig.~\ref{f-sigpint},
and replacing the true quantities $r_v$ and $\sigma_a$, with their
estimates $\tilde{r}_v$ and $\sqrt{3} \sigma_{a,p}$, respectively, we
obtain an estimate of $\sigma_v$ which we use to determine the
$M_{\sigma}$ estimate of $M_v$ through eq.~\ref{m-sig}.

In order to verify the performance of these two mass estimators, we
have applied them to the clusters in the simulation, making use of the
{\em full} phase-space information. In this case, there is no need to
apply the interloper-rejection procedure outlined above (step 2),
since we randomly select 5000 DM particles in each cluster within a
{\em sphere} (not a cylinder) of radius $a \equiv r_v$ (when this
sphere contains less than 5000 DM particles, all particles are
selected). Also the de-projection factor $3 \pi/2$ (step 6) is not
needed. As for the estimate of $M_{\sigma}$, we directly apply
eq.~\ref{m-sig}.  In Fig.~\ref{f-dm5000} we show the ratio between the
virial ($\tilde{M}_v$) and the true masses (upper panel), and
$M_{\sigma}$ and the true masses (lower panel) for our sample of 62
clusters.  On average we find $\tilde{M}_v/M_v=1.03$ with a dispersion
of 0.09, and $M_{\sigma}/M_v=1.00$ with a dispersion of 0.12 (see also
Tormen et al. \cite{torm97} for a similar analysis). Perfect identity
is not expected, since clusters are not fully virialized, but the good
agreement indicates that the estimators are unbiased and the clusters
in the simulation are close to virialization within $r_v$, as expected
for $z \sim 0$ clusters in a $\Lambda$CDM cosmology.

\section{Dynamical analysis in projected phase-space}\label{s-2d}
In order to mimic observations, the simulated clusters are placed at a
distance of 250~$h^{-1}$~Mpc from the hypothetical observer. Three
orthogonal projections are considered for each cluster, leading to a
total of 186 ($62 \times 3$) cluster projections (except when we
remove the cases with significant evidence for subclustering, see
Sect.~\ref{s-sub}). DM particles or galaxies are then selected within
cylinders of given radius, chosen to be a fraction (from 1/3 to 1) of
an Abell radius. This is meant to reproduce typical observational
procedures where the virial radius of the system is unknown prior to
observation. The length of the cylinder is set by the simulation cube.

In the following we analyse how cluster mass estimates are affected by
several observational effects. In particular, we consider the effects
of different sample sizes, different observational apertures,
observational errors, incompleteness, and subclusters detected in
projected phase-space.  In principle, it would be desirable to perform
all these analyses using the simulated galaxies as tracers of the
potential, rather than the DM particles. As a matter of fact, the
analyses performed on the simulated galaxies are likely to provide a
reliable picture of the real observational situation, since the
phase-space distributions of simulated galaxies are not too different
from the observed ones (see Sect.~\ref{s-sims}). On the other hand,
the DM particles are not distributed like the galaxies in the
simulated clusters (see Sect.~\ref{s-gal}).  Unfortunately, however,
there are only few galaxies per cluster in our simulated set, hence
small-number statistical noise becomes a dominant source of
scatter. In order to overcome this problem, we have also considered
DM particles as tracers of the gravitational potential, even if,
of course, these are not true 'observables'.

The most relevant results of the dynamical analyses in projected
phase-space are summarized in Table~\ref{t-pratios}. The different
analyses performed are identified by their number in the 1st Col. of
Table~\ref{t-pratios}. Col.~2 lists the observational aperture size,
$a$, in $h^{-1}$ Mpc, i.e. the radius of the cylinder within which
particles or galaxies are selected.  Col.~3 lists the number of DM
particles per projection initially selected for the analysis within a
cylinder of given radius. In the case of galaxies, all galaxies (of
the specified type) are always selected, and this number changes from
projection to projection, hence we list the average over all
projections.  This is also the case for DM particles, when we select
an identical number of DM particles as of available galaxies within
the cylinder (see Sect.~\ref{s-gal}). In Col.~4 'DM' or 'G' indicates
whether DM particles or Galaxies are chosen as tracers of the
potential.  When only galaxies of a given type are selected, this is
further specified after the letter 'G'.  Col.~5 lists the type of
selection performed on DM particles, which can either be random
('Rdm') or of a type that reproduces a single-mask, multi-slit
observation ('Slit'). In the case of galaxies, 'All' indicates that
all galaxies in the cylinder are selected.  The note 'projections with
subclusters excluded' indicates that the definition of the sample used
is the same as in the line above, except that cluster projections with
significant evidence of substructure have been removed from the
sample.  Col.~6 and 7 list the average of the number of DM particles
or galaxies selected as cluster members within the chosen aperture,
and the fraction $f_{out}$ of these that actually lie {\em outside the
sphere} of radius equal to the chosen aperture, and that we call
``interlopers''.  Cols.~8 and 9 list the average ratios of the
estimated-to-true velocity dispersion $\sqrt{3}
\sigma_{a,p}/\sigma_a$, and of the estimated-to-true harmonic mean
radius $\frac{\pi}{2} r_{a,p}/r_a$, respectively, all quantities being
estimated within the given aperture $a$ (note that the projected
quantities are multiplied by the deprojection factors).  Col.~10 lists
the average ratio of the estimated-to-true virial radius,
$\tilde{r}_v/r_v$.  Cols.~11 and 12 list the average ratios of the two
mass estimates (the virial estimate and the velocity-dispersion based
estimate, respectively) to the true mass, $\tilde{M}_v/M_v$ and
$M_{\sigma}/M_v$, respectively. The dispersions of the quantities
listed in Cols. 6--12 are given (within brackets) in the line
immediately below. Cols.~13 and 14 list the fractions of cluster
projections for which the virial mass is a strong under- or
over-estimate of the true mass, $\tilde{M}_v/M_v < 1/2$, and
$\tilde{M}_v/M_v > 2$, respectively.

\begin{table*}
\centering
\caption[]{The ratios between the estimated dynamical quantities in projection
and the true ones. Quantities in brackets are the dispersions on the
estimated quantities.}
\label{t-pratios}
\begin{tabular}{rcrccrcccccccc}
\hline
Id. & $a$ & $N$ & Tracer & Sel. &
$N_m$ & $f_{out}$ & $\sqrt{3} \sigma_{a,p}/\sigma_a$ & $\frac{\pi}{2} r_{a,p}/r_a$ &
  $\tilde{r}_v/r_v$ & $\tilde{M}_v/M_v$ & $M_{\sigma}/M_v$ & $f_{<0.5}$ & $f_{>2}$ \\
   & ($h^{-1}$ Mpc)   &   & &  & & & & & & & & \\
\hline
 1 &1.50 &500 & DM &  Rdm &  402  & 0.18 & 0.93   & 1.22   & 1.08   & 1.10   & 0.88   & 0.01 & 0.03 \\
   & &     &      &     & (48) & (0.11) & (0.10) & (0.12) & (0.10) & (0.31) & (0.28) &  & \\
 2 & \multicolumn{4}{c}{projections with subclusters excluded} & 396 & 0.17 & 0.90 & 1.21 & 1.06 & 1.02  & 0.81 & 0.00 & 0.00 \\
   & &     &      &     & (50) & (0.10) & (0.10) & (0.11) & (0.08) & (0.25) & (0.24) &  & \\
 3 &1.50 &100 & DM &  Rdm &   80  & 0.18 & 0.93   & 1.20   & 1.07   & 1.10   & 0.81   & 0.04 & 0.10 \\
   & &     &      &     &  (9)  & (0.11) & (0.14) & (0.19) & (0.14) & (0.46) & (0.38) &  & \\
 4 &1.50 & 20 & DM &  Rdm &   17  & 0.20 & 0.89   & 1.37   & 1.11   & 1.25   & 0.83   & 0.09 & 0.25 \\
   & &     &      &     &  (2) & (0.14) & (0.27) & (0.28) & (0.25) & (0.79) & (0.58) &  & \\
 5 &1.50 &500 & DM &  Slit &  198  & 0.26 & 0.90   & 1.63   & 1.18   & 1.51   & 0.83   & 0.00 & 0.16 \\
   & &     &      &     & (51) & (0.13) & (0.12) & (0.17) & (0.10) & (0.41) & (0.28) &  & \\
 6 &0.50 &500 & DM &  Rdm &  459  & 0.28 & 0.94   & 1.20   & 0.97   & 1.08   & 0.87   & 0.01 & 0.02 \\
   & &     &      &     & (46) & (0.09) & (0.10) & (0.07) & (0.09) & (0.32) & (0.29) &  & \\
 7 &1.50 &39 &  G  & All &   30  & 0.26 & 0.95   & 1.39   & 1.16   & 1.40   & 0.94   & 0.11 & 0.27 \\
   & &     &      &     & (12) & (0.15) & (0.27) & (0.25) & (0.24) & (0.85) & (0.62) &  & \\
 8 &1.50 &39 &  DM & Rdm &   29  & 0.24 & 0.98   & 1.37   & 1.17   & 1.46   & 1.03   & 0.03 & 0.31 \\
   & &     &      &     &  (12)  & (0.14) & (0.26) & (0.26) & (0.24) & (0.89) & (0.68) &  & \\
 9 &1.50 & 19 &  G $z_f \geq 1.25$ & All &   16  & 0.18 & 0.89   & 1.21   & 1.04   & 1.01   & 0.80   & 0.20 & 0.15 \\
   & &     &      &     &  (8) & (0.15) & (0.30) & (0.30) & (0.26) & (0.74) & (0.64) &  & \\
10 &1.50 & 19 & DM &  Rdm &   15 & 0.25 & 0.93 & 1.41   & 1.13   & 1.32   & 0.92   & 0.14 & 0.34 \\
   & &     &      &     &  (7)  & (0.18) & (0.37) & (0.32) & (0.35) & (1.14) & (0.83) &  & \\
\hline
\end{tabular}
\end{table*}

\subsection{Projection effects}\label{s-proj}
In order to quantify the importance of projection effects on cluster
mass estimates, we first consider a large number (500) of reliable
tracers of the gravitational potential (DM particles), with no errors
on their velocities, positions, and no radial incompleteness bias (see
Table~\ref{t-pratios}, Id. no.1). The 500 DM particles are randomly
selected along each of three orthogonal line-of-sights for each
cluster, within a cylinder of 1.5 $h^{-1}$~Mpc radius. Interlopers are
then rejected using the procedure described in Sect.~\ref{s-mass}. The
case studied here is of course an idealized situation never achieved
observationally, but this analysis serves to emphasize the importance
of projection effects on cluster mass estimates, independently of any
other observational bias.

\begin{figure}
\centering
\resizebox{\hsize}{!}{\includegraphics{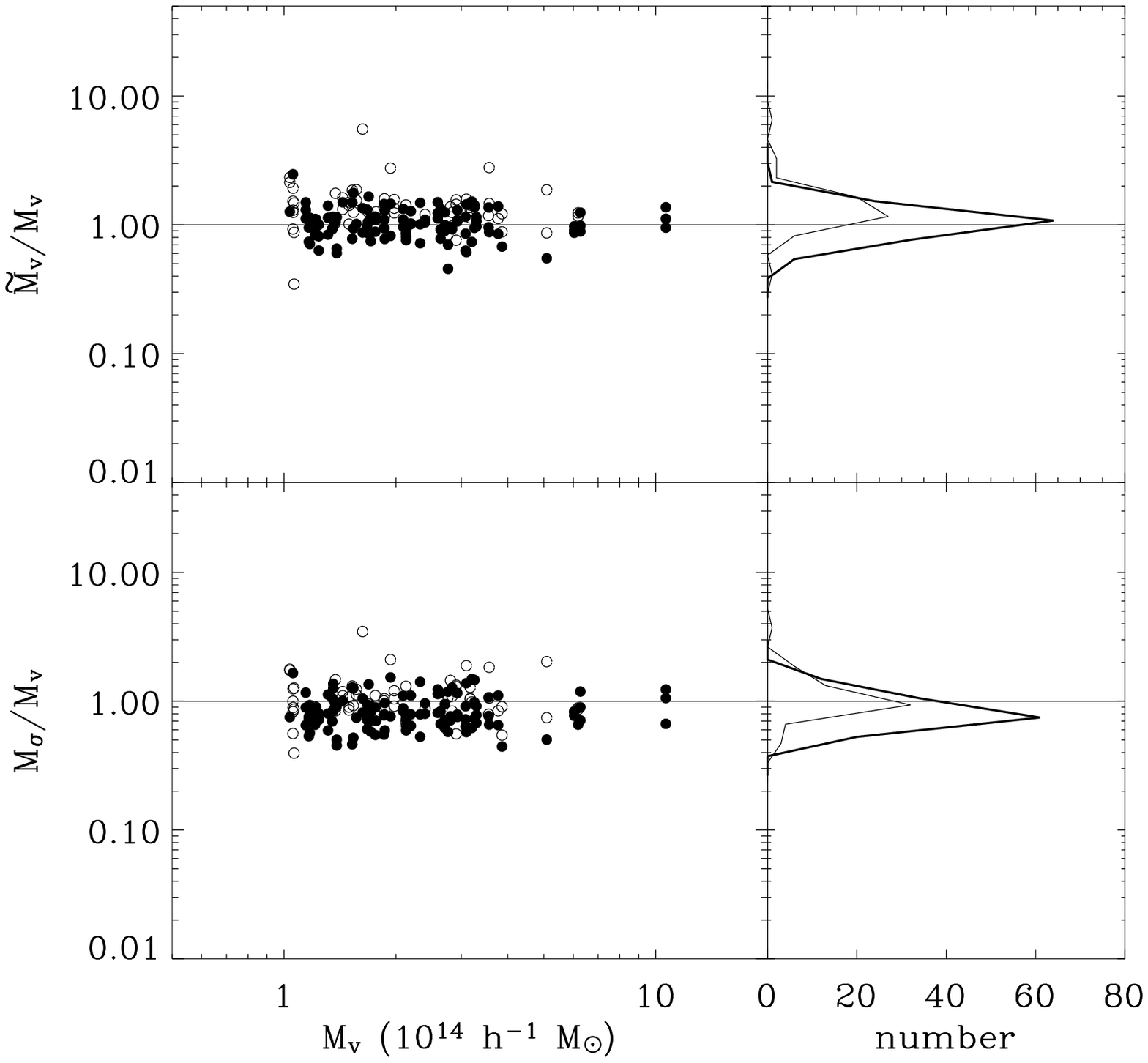}}
\caption{The comparison between the masses estimated using only
projected phase-space information and the true cluster masses.
Top-left panel: the ratio between the virial and the true mass,
$\tilde{M}_v/M_v$ vs. the true mass $M_v$. Filled dots identify those
projections where no significant evidence for substructure is found at
the 0.01 c.l. Top-right panel: distributions of $\tilde{M}_v/M_v$ for
the projections with and without significant evidence for substructure
(thin and thick line, respectively).  Bottom panels: same as top
panels, but for $M_{\sigma}$ in lieu of $\tilde{M}_v$. The masses
$\tilde{M}_v$ and $M_{\sigma}$ are estimated from projected
distributions of DM particles. Initially, 500 DM particles are
randomly selected along three orthogonal line-of-sights in each
cluster, within a cylinder of 1.5 $h^{-1}$Mpc radius. Interlopers are
then rejected using the procedure described in
Sect.~\ref{s-mass}. Note that the vertical range is different from
that of Fig.~\ref{f-dm5000} and the vertical axis is now logarithmic.}
\label{f-dmp500}
\end{figure}

On average, both the virial mass estimate and the $M_{\sigma}$
estimate are accurate to $\simeq 10$\%, but the scatter around the true
mass values for the different projections is quite large (see
Fig.~\ref{f-dmp500}).  The virial estimate somewhat overestimates the
true mass, and this is caused by an overestimate of the harmonic mean
radius, while $M_{\sigma}$ somewhat underestimates the true mass, and
this is caused by an underestimate of the velocity dispersion (see
Table~\ref{t-pratios}).

In order to better understand the effects of projection, we
examine the projected phase-space distribution of a stacked cluster in
some detail. The stacked cluster is built as follows. In each cluster
of the simulation set we randomly select 500 DM particles along three
orthogonal projections, and identify the cluster members.
We then normalize the clustercentric distances of
these particles by the virial radius of the cluster they belong to,
and their velocities (relative to their cluster mean velocity) by the
velocity dispersion of the cluster they belong to, computed within one
virial radius. We then stack the 62 clusters using the normalized
radii and velocities. Finally, we return to physical units by
multiplying the normalized radii and velocities of the DM particles in
the stacked cluster, respectively by the average virial radius (0.99
h$^{-1}$ Mpc), and by the average velocity dispersion (657
km~s$^{-1}$) of the 62 clusters.

\begin{figure}
\centering \resizebox{\hsize}{!}{\includegraphics{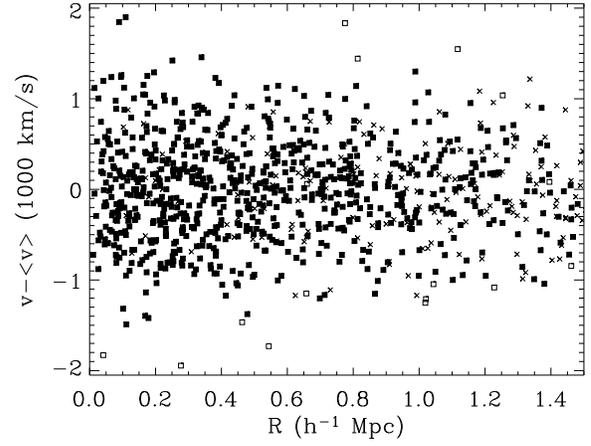}}
\caption{Line-of-sight velocity vs. projected clustercentric distance
for a subset of the DM particles in the stacked cluster (see
text). Filled squares indicate DM particles contained within a sphere
of 1.5 $h^{-1}$~Mpc radius and selected as cluster members by the
interloper rejection procedure, open squares indicate DM particles
located within this same sphere yet incorrectly rejected as
interlopers, and x symbols indicate DM particles located outside this
same sphere and yet selected as cluster members (interlopers).}
\label{f-rvall}
\end{figure}

In Fig.~\ref{f-rvall} we show the projected phase-space distribution
of a randomly selected subset of the DM particles in the stacked
cluster. Three orthogonal projections are stacked together in the
Figure.  Different symbols are used to represent: i) DM particles
contained within a sphere of 1.5 $h^{-1}$~Mpc radius and selected as
cluster members by the interloper rejection procedure (filled
squares), ii) DM particles located within this same sphere yet
incorrectly rejected as interlopers (open squares), and iii) DM
particles located outside this same sphere and yet selected as cluster
members (x symbols). Clearly, the interloper rejection technique
excludes only a few DM particles that are in fact within the 1.5
$h^{-1}$~Mpc sphere, but keeps several DM particles that are outside
the 1.5 $h^{-1}$~Mpc sphere but have projected velocities that make
them indistinguishable from the real cluster members in the projected
phase-space diagram (what we have called interlopers, see
Sect.~\ref{s-2d}).

\begin{figure}
\centering \resizebox{\hsize}{!}{\includegraphics{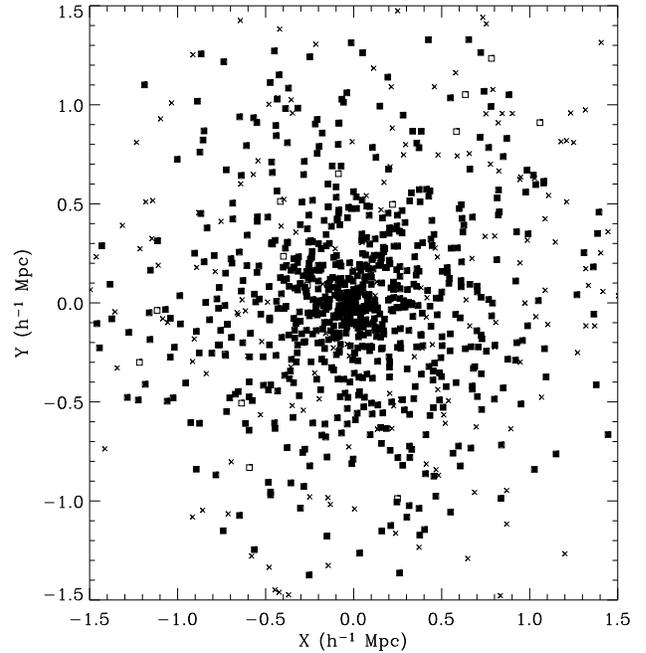}}
\caption{The projected spatial distribution of the same DM
particles shown in Fig.~\ref{f-rvall}. Symbols are the same as in
Fig.~\ref{f-rvall}.}
\label{f-xyall}
\end{figure}

The underestimate of the velocity dispersion of the cluster is partly
due to the rejection of real cluster members at relatively high
velocities. Most of the underestimate is however due to the inclusion
of interlopers that are currently infalling toward the cluster along a
filament.  These interlopers lie within $\la 10$ $h^{-1}$~Mpc from the
cluster centre (more distant DM particles have quite a different
recession velocity from the cluster mean and are rejected by the
interloper technique) and their infall motion makes their velocities
resemble the average cluster velocity. In addition, they have an
intrinsically small velocity dispersion within their filament. As a
consequence, the projected velocity dispersion of these interlopers
turn out to be smaller than that of the cluster within the selected
aperture. On the other hand, their spatial distribution is not as
centrally concentrated as that of the cluster members within the
selected aperture, as shown in Fig.~\ref{f-xyall}.  This causes the
harmonic mean radius to be overestimated.

Therefore, interlopers cause the overestimate of the harmonic mean
radius, and, at the same time, the underestimate of the velocity
dispersion, as already noted by C97 and Diaferio et
al. (\cite{diaf99}). Since the strongest of the two effects is the
harmonic mean radius overestimate, also the virial mass is
overestimated. On the other hand, $M_{\sigma}$ provides an
underestimate of the mass because it only depends on the velocity
dispersion estimate.

\begin{figure}
\centering \resizebox{\hsize}{!}{\includegraphics{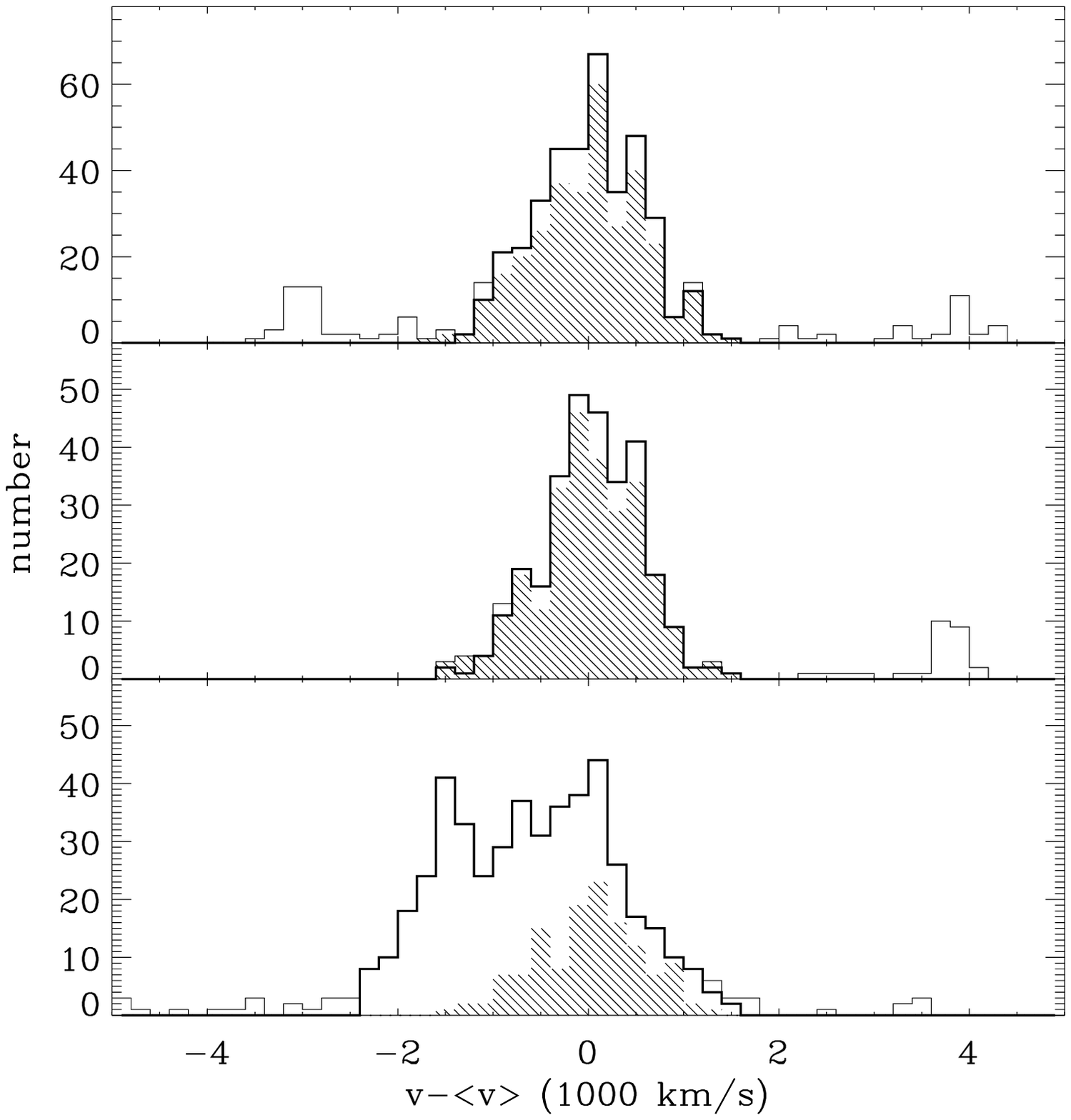}}
\caption{The velocity distributions of a random subset of DM particles
of a simulated cluster, along three orthogonal projections. Thin lines
indicate the velocity distributions of all DM particles within a
cylinder of radius 1.5 $h^{-1}$~Mpc, thick lines the velocity
distributions of those DM particles identified as cluster members, and
the hatched histograms show the velocity distributions of the DM
particles located within a {\em sphere} of 1.5 $h^{-1}$~Mpc radius.}
\label{f-vh4613}
\end{figure}

Despite these problems, the interloper selection technique seems to
work reasonably well. On average only $\sim 1$\% of the real cluster
members are rejected as interlopers, while on average 18\% of the
selected members are unrecognized interlopers (see $f_{out}$ in
Table~\ref{t-pratios}). However, projection effects can become
critical for individual cases. In 3\% of the cases the cluster masses
are overestimated by a factor $>2$, and in 1\% of the cases they are
underestimated by a factor $<1/2$, using the virial mass estimator
(see Fig.~\ref{f-dmp500}, and the last two columns in
Table~\ref{t-pratios}).  Large errors in the virial mass estimates
correspond to extreme cases of failure of the interloper rejection
algorithm, when either a compact dynamical system is artificially
split in two by the weighted gap algorithm, or, more frequently, a
physically distinct group is merged together with the main cluster.
This last case is illustrated in Fig.~\ref{f-vh4613}, where we plot
the velocity distributions of i) the DM particles within a sphere of
1.5 $h^{-1}$~Mpc radius (hatched histogram), ii) all DM particles
within the cylinder of same radius (thin line), and iii) the DM
particles identified as cluster members (thick line). In one of the
three projections (bottom panel of Fig.~\ref{f-vh4613}) the cluster
velocity distribution is heavily contaminated by that of a foreground
group, and the interloper rejection technique fails to isolate the
cluster members. Note, however, that large errors in the mass
estimates generally occur only for one of the three orthogonal
projections considered per cluster.

\subsection{Subclustering}\label{s-sub}
The presence of subclustering has long been thought to potentially
cause incorrect cluster mass estimates (e.g. Bird \cite{bird95}).
Several algorithms have been developed over the years to address this
problem (see, e.g., Girardi \& Biviano \cite{gira02}, and references
therein).  One of the most commonly used (and easy to implement)
algorithms is that of Dressler \& Shectman (\cite{dres88}; DS
hereafter). Here we apply this algorithm (or, actually, Biviano et
al.'s \cite{bivi02} version of it) to the simulated clusters seen
along each of three orthogonal projections. As before (see
Sect.~\ref{s-proj}) 500 DM particles are randomly selected within a
cylinder of 1.5 $h^{-1}$~Mpc radius; the DS test is applied on
those particles that are selected as cluster members.

In 59 cases out of 186 the DS test provides a probability $\geq 0.99$
that the cluster, as seen along the chosen projection, has significant
subclustering. The cluster projections with $\geq 0.99$ subclustering
probability are also those with the largest mass overestimates. As an
example, in the case-study illustrated in Fig.~\ref{f-vh4613},
subclustering is detected with high significance along the projection
shown in the bottom panel.  Among the 127 cluster projections {\em
without} evidence for subclustering, only one has its virial mass
overestimated by a factor $> 2$. As a consequence, the virial mass
estimates are now closer to the true cluster masses, on average (see
Fig.~\ref{f-dmp500} and Table~\ref{t-pratios}, Id. no.2).

Hence the simple DS test appears to be a useful tool for identifying
those cases of large virial mass overestimates. However, the situation
is not so simple. When only 100 DM particles are selected instead of
500, the DS test gives a probability $\geq 0.99$ only for 28 out of
the 186 cluster projections. Clearly, when the sample size is reduced
the DS statistics is less significant. Yet, observational results
based on even smaller sample sizes, suggest a higher fraction of
clusters with subclustering, $\sim 1/3$ (e.g. DS; Biviano et
al. \cite{bivi97}). Hence the simulated clusters seem to display a
lower amount of subclustering, on average, than real clusters.

\subsection{Undersampling}\label{s-under}
In order to study how the accuracy of mass estimates depends on the
number of cluster members available, we randomly select from 10 to 500
DM particles within cylinders of 1.5 $h^{-1}$~Mpc radius, along three
orthogonal projections. Cluster members are then selected using the
interloper rejection procedure described in Sect.~\ref{s-mass}.

\begin{figure}
\centering
\resizebox{\hsize}{!}{\includegraphics{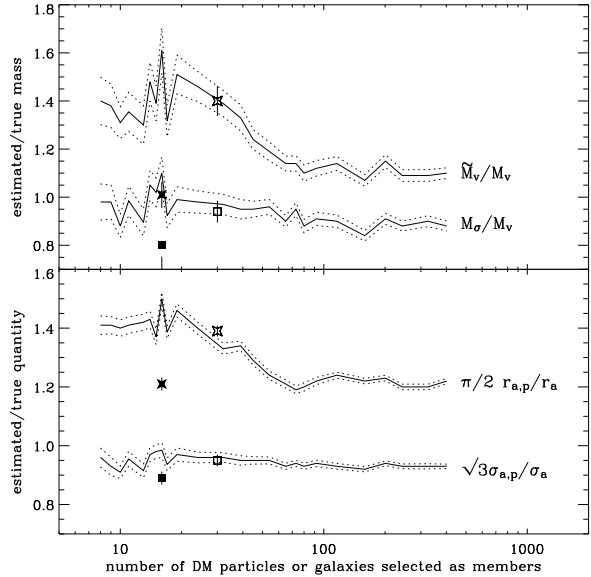}}
\caption{The ratio of estimated to true dynamical quantities, averaged
over all cluster projections, as a function of the number of DM
particles selected as cluster members. Upper panel: $\tilde{M}_v/M_v$
(lines above) and $M_{\sigma}/M_v$ (lines below).  Lower panel: $\frac{\pi}{2}
r_{a,p}/r_a$ (lines above) and $\sqrt{3} \sigma_{a,p}/\sigma_a$ (lines
below). Dotted lines represent 1-$\sigma$ c.l. on the averages. The
symbols represent the values obtained using galaxies (rather than DM
particles) as tracers. Empty symbols are for all galaxies; filled
symbols for galaxies with $z_f \geq 1.25$ only. Stars refer to
$\tilde{M}_v/M_v$ and $\frac{\pi}{2} R_h/r_h$, squares to $M_{\sigma}/M_v$ and
$\sigma_{200}/\sigma_v$.}
\label{f-ssize}
\end{figure}

Reducing the sample size increases the scatter of the dynamical
estimates and affects the virial mass estimates (see
Fig.~\ref{f-ssize} and Table~\ref{t-pratios}, Id. nos. 1, 3, and
4). The average $r_{a,p}/r_a$ ratio tends to increase with decreasing
number of selected cluster members, $N_m$, for $N_m \la 60$, until it
reaches a plateau for $N_m \la 20$. On the other hand, the average
$\sqrt{3} \sigma_{a,p}/\sigma_a$ ratio hardly changes with the size of
the sample.  As a consequence, for samples of $\la 60$ cluster
members, the simpler $M_{\sigma}$ estimate appears to be a better
predictor of the true mass than the virial mass estimate,
$\tilde{M}_v$, which is affected by the bias of $r_{a,p}$.

The increasing overestimate of the harmonic mean radius (and hence of
the virial mass) as the size of the sample decreases is mostly due to
an increasing fraction of interlopers, from 18\% for $N_m \ga 60$, to
23--25\% for $N_m \la 40$. Part of the effect is however induced by a
statistical bias that affects the harmonic mean radius estimate for
small sample sizes. This can be seen by computing the harmonic mean
radius in projection, but using only real cluster members.  We find
that in this case, the ratio $r_{a,p}/r_a$ remains close to unity for
$N_m \ga 60$, then starts increasing with decreasing $N_m$, until it
becomes $\simeq 1.1$ for $N_m \simeq 20$.

\subsection{Radial-dependent incompleteness}\label{s-slit}
When a cluster field is observed in spectroscopy, completeness to a
given limiting magnitude is rarely achieved. However, apart from the
very bright cluster galaxies, there is no evidence for luminosity
segregation in galaxy clusters (e.g. Biviano et al. \cite{bivi92},
\cite{bivi02}; Adami et al. \cite{adam98a}; Cappi et al. \cite{capp03};
Goto \cite{goto05}; Yang et al. \cite{yang05}). Hence a random
selection of the available galaxies in a cluster field, even if not
complete down to a given magnitude, should produce a sample with an
unbiased projected phase-space distribution. In the previous analyses
(Sects.~\ref{s-proj}, \ref{s-sub}, \ref{s-under}) such a random
selection was indeed applied. However, this is not always an easy task
to accomplish observationally. In fact, the number of slits (fibers)
per mask in multi-slit (multi-fiber) spectrographs is fixed, hence the
central, high-density regions of galaxy clusters are often sampled to
a brighter magnitude than the external regions. As a consequence, the
spatial distribution of the galaxies selected for spectroscopy turns
out to be more extended than the parent spatial distribution of
cluster galaxies. Said otherwise, a higher fraction of galaxies is
selected from the total in the outer regions, i.e. the incompleteness
is not random but depends on radius.

Ignoring the problem of radial-dependent incompleteness can have quite
a catastrophic effect on the cluster mass estimate, as we show in the
following. We simulate an observational set-up in which our clusters
are observed with a multi-slit spectrograph with four quadrants, each
13 arcmin on a side, with a typical slit separation of 10 arcsec (we
remind the reader that the clusters have been set at a distance of
$250$ $h^{-1}$~Mpc from the observer). The results are listed in
Table~\ref{t-pratios}, Id. no.5. In comparing these results to those
obtained in the case of a random selection of tracers
(Table~\ref{t-pratios}, Id. no.1), we note three differences.  First,
the average number of selected DM particles is only 198 out of the
initial 500 randomly chosen in a cylinder of 1.5 $h^{-1}$~Mpc
radius. Second, the fraction of interlopers is increased (26\%
vs. 18\%).  Third, the harmonic mean radius is strongly
overestimated in projection.  All these differences are due to the
geometrical constraints imposed by the multi-slit single-mask
observation, that forces a sparser sampling of the dense cluster core
relatively to the cluster outskirts. The observational spatial
distribution of the selected tracers is thus less centrally
concentrated than the underlying parent distribution.

In order to avoid this bias, the central cluster regions must be
observed with more masks per unit area than the external cluster
regions. Alternatively, photometric observations can be used to
estimate, and correct for, the radial incompleteness of the
spectroscopic sampling. When neither option is viable, it is better to
rely on the $M_{\sigma}$ estimate, which is much less affected by
problems of incompleteness since cluster velocity dispersion profiles
have only a mild radial dependence (see, e.g., Girardi et
al. \cite{gira96} and Fig.~\ref{f-sigpint}).

\subsection{Other observational effects}\label{s-others}
Other observational effects that could, in principle, affect a cluster
mass determination, are i) a different size of the observational
aperture (the radius of the cylinder within which the tracers are
selected), ii) the uncertainty on the determination of the cluster
centre, and iii) errors on the redshifts of cluster galaxies. All
these effects can change the projected phase-space distributions of
the tracers, thereby affecting in principle also the identification of
cluster members.

In order to estimate the effect that different aperture sizes have on
the cluster mass estimates, we compare the results obtained by
selecting DM particles within cylinders of 1.5, 0.75, and 0.5
$h^{-1}$~Mpc radii. As the aperture $a$ of the cylinder is decreased,
more particles are selected as cluster members out of the initial 500
distributed in the cylinder, but a higher fraction of the selected
members are actually located outside a sphere of radius $a$. As a
consequence, the results of the dynamical analysis are similar for
different apertures (compare the results for an aperture of 0.5
$h^{-1}$~Mpc with those for an aperture of 1.5 $h^{-1}$~Mpc,
Id. nos. 6 and 1, respectively, in Table~\ref{t-pratios}).

In order to simulate the observational uncertainty in the
determination of a cluster centre, a random offset is added to the
position of the real cluster centre (see Sect.~\ref{s-mass}). The
offset is randomly taken from a lognormal distribution with average of
40~kpc and dispersion of 50~kpc, modeled after the observed
distribution of uncertainties in the centre positions of ENACS
clusters (Adami et al. \cite{adam98b}). We find that centering errors of this
size have essentially no effect on the accuracy of mass estimation.

Velocity errors contribute to increase the estimate of the cluster
velocity dispersion. In order to simulate the observational errors on
galaxy redshifts, a random velocity offset is added to the real DM
particle velocity. The offset is randomly taken from a gaussian
distribution with width of 70 or 210 km~s$^{-1}$. An error of 70
km~s$^{-1}$ is typical for observations of nearby clusters
observations (see, e.g., Katgert et al. \cite{katg96}), while an error
three times larger corresponds to the observational uncertainties of
galaxy redshifts in distant clusters (e.g. Demarco et
al. \cite{dema05}). As long as the velocity dispersion estimate is
corrected following the prescriptions of Danese et
al. (\cite{dane80}), the dynamical estimates are not significantly
affected by errors on the velocities of the tracer particles. Note
however that the 62 clusters used in our analysis have a median
line-of-sight velocity dispersion of 622 km~s$^{-1}$. Galaxy velocity
errors are likely to become more important for mass estimates of
low-velocity dispersion groups.

\subsection{Galaxies}\label{s-gal}
\begin{figure}
\centering
\resizebox{\hsize}{!}{\includegraphics{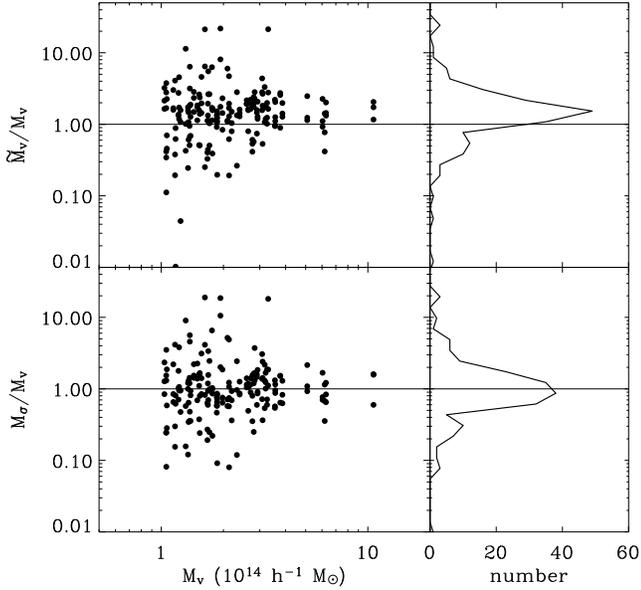}}
\caption{Same as Fig.~\ref{f-dmp500} but for galaxies instead of DM
particles. No distinction is made here between projections with
and without detection of substructures.}
\label{f-gals32p}
\end{figure}

The limited number of simulated galaxies per cluster makes it
impossible to study in detail all kinds of different observational
effects, as it was done for the DM particles. But galaxies have the
advantage that early- and late-types can be distinguished, based on
their formation redshift $z_f$ (see Sect.~\ref{s-sims}). We start by
considering all available galaxies within a cylinder of
1.5~$h^{-1}$~Mpc radius. The results are shown in
Table~\ref{t-pratios}, Id. no. 7 and Fig.~\ref{f-gals32p}. Dynamical
estimates based on galaxies are significantly affected by the presence
of interlopers. About 1/4 of the selected cluster members are in fact
outside the sphere of 1.5~$h^{-1}$~Mpc radius. As a consequence, both
$r_{a,p}$ and $\tilde{M}_v$ overestimate the harmonic mean radius and
the mass, respectively, by about 40\%. In about 1/3 of the
projections, the virial mass estimates are wrong by a factor of two or
greater. On the other hand, the $\sigma_{a,p}$ and $M_{\sigma}$
estimates are (on average) almost correct (to within 5--6\%).

\begin{figure}
\centering
\resizebox{\hsize}{!}{\includegraphics{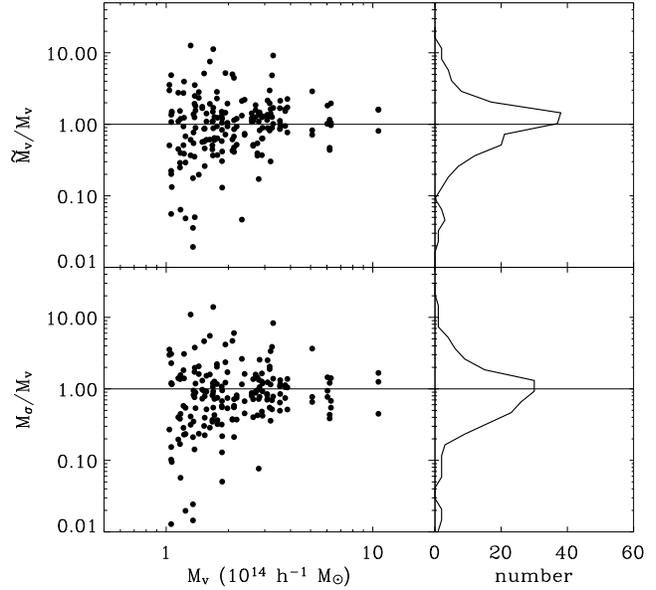}}
\caption{Same as Fig.~\ref{f-gals32p} but only for galaxies with
$z_f \geq 1.25$.}
\label{f-gals32p_zf125}
\end{figure}

In order to compare these results with those obtained using DM
particles as tracers of the potential, we randomly select within each
cylinder of 1.5~$h^{-1}$~Mpc radius a number of DM particles identical
to that of the galaxies in that same cylinder.  In this way we make
sure that the effects of undersampling (see Sect.~\ref{s-under}) are
the same for the samples of galaxies and DM particles. The results are
given in Table~\ref{t-pratios}, Id. no. 8. The differences are
very marginal.

When only early-type galaxies (i.e. those with $z_f \geq 1.25$) are
selected, the results of the dynamical analysis are quite different
from the case in which all galaxies are selected (compare Id. nos. 9
and 7 in Table~\ref{t-pratios} respectively; see also
Fig.~\ref{f-gals32p_zf125}). In particular, the average fraction of
interlopers is now reduced to 18\%, and the average value of
$\tilde{M}_v$ is now almost identical to the real mass value.  On the
other hand, the $M_{\sigma}$ estimate is 20\% too low, as a
consequence of a dynamical segregation effect (see the discussion
below).  The number of clusters with a mass estimate wrong by a factor
of at least two is still high, about 1/3 of the total, but this is
expected because of the large scatter in the mass estimates due to the
small number of available tracers.

As before, we compare these results with those obtained by selecting
the same number of DM particles as early-type galaxies, within the
cylinder of 1.5~$h^{-1}$~Mpc radius. The results are listed in
Table~\ref{t-pratios}, Id. no. 10, and are quite different from those
obtained using early-type galaxies, since, in particular, the fraction
of interlopers is higher, and $\tilde{M}_v$ significantly overestimate
the true mass.

\begin{figure}
\centering
\resizebox{\hsize}{!}{\includegraphics{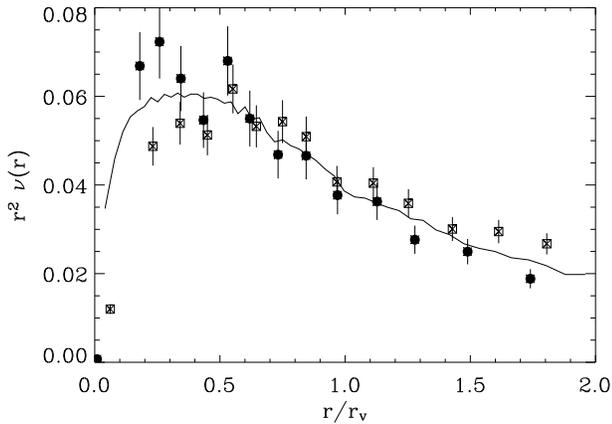}}
\caption{The 3-d number density profiles of DM particles (solid line), 
galaxies with $z_f \geq 1.25$ (dots), and all galaxies 
(squared x's) in the stacked cluster.}
\label{f-nu}
\end{figure}

\begin{figure}
\centering
\resizebox{\hsize}{!}{\includegraphics{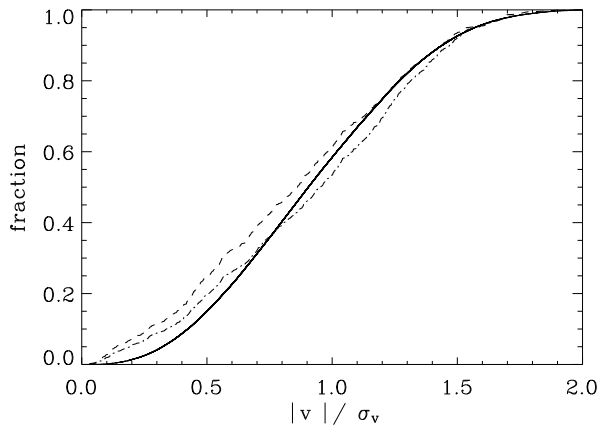}}
\caption{The cumulative distributions of the 3-d velocities of DM
particles (solid line), galaxies with $z_f \geq 1.25$ (dashed line),
and all galaxies (dash-dotted line) in the stacked cluster.}
\label{f-vdist}
\end{figure}

In order to understand the differences found between the mass
estimates obtained using DM particles, all galaxies, and early-type
galaxies, we consider their relative distributions in {\em full}
phase-space. For this we build a stacked cluster, as in
Sect.~\ref{s-sims}.  The 3-d number density profiles of DM particles
and galaxies in the stacked cluster are shown in
Fig.~\ref{f-nu}. Galaxies are clearly less centrally concentrated than
DM particles. This is confirmed by a Rank-Sum (RS hereafter)
statistical test that reject the null hypothesis that the DM particles
and the galaxies have the same radial distribution with a high
confidence level (c.l. in the following), $>0.999$ (see also Nagai \&
Kravtsov \cite{naga05}).  Interestingly, this is in agreement with the
observations that show a decreasing mass-to-light ratio as a function
of clustercentric distance (e.g. Biviano \& Girardi \cite{bivi03};
Rines et al. \cite{rine04}).  On the other hand, the distribution of
the normalised velocities, $\mid v \mid / \sigma_v$, of galaxies is
not significantly different from that of DM particles (RS test
c.l. 0.829; see Fig.~\ref{f-vdist}).

The spatial distribution of early-type galaxies is significantly more
concentrated than that of all galaxies (RS-test c.l. $>0.999$), and
marginally more concentrated than that of DM particles (RS-test
c.l. 0.939; see Fig.~\ref{f-nu}). The velocity distribution of
early-type galaxies is significantly different from those of all the
galaxies and the DM particles (RS-test c.l. $>0.999$; see
Fig.~\ref{f-vdist}). The high significance of this difference is a
consequence of the large size of the data-sets, and does not reflect a
large velocity bias: on average the modulus of the velocities of
early-type galaxies is 0.94 that of DM particles and 0.93 that of all
the galaxies. In real clusters, the different velocity
distribution of early- and late-type galaxies is a well established
observational fact (Moss \& Dickens \cite{moss77}; Sodr\'e et
al. \cite{sodr89}). The factor by which the velocity dispersion of
early-type galaxies is lower than the velocity dispersion of all the
cluster galaxies, is 0.93, on average, for the ENACS clusters
(as can be derived from the estimates given in Biviano et
al. \cite{bivi97}), a value identical to the one found here for
simulated galaxies.

The different spatial distributions of all galaxies and DM particles
does not seem to influence the results of the dynamical analysis in
projected phase-space. One would expect the wider spatial distribution
of galaxies to result in a larger $r_a$, and hence $r_{a,p}$ estimate.
However, the $r_{a,p}$ estimate is dominated by the presence of
interlopers. Since the fractions of interlopers among galaxies and
among DM particles are similar, also the $r_{a,p}$ estimates are
similar.

When only early-type galaxies are selected, the fraction of
interlopers decreases because the fraction of early-type galaxies {\em
outside} clusters is low, 29\%, much lower than the fraction of
early-type galaxies {\em within} clusters (see Sect.~\ref{s-sims}). As
a consequence, also the cluster $r_{a,p}$ estimate is closer to the
real $r_a$ value, than in the case of DM particles, although still
$\simeq 20$\% too high. The $r_a$ overestimate does not result in an
overestimate of the virial mass because $\sigma_{a,p}$ underestimates
the real $\sigma_a$ by $\simeq 10$\%. This underestimate is caused by
the narrower velocity distribution of early-type galaxies relatively
to DM particles.

In conclusion, galaxies have a biased distribution relatively to DM
particles. Considering all the cluster galaxies, they have a wider
spatial distribution than DM particles, but a similar velocity
distribution. Considering instead only ``early--type'' galaxies, these
have both a narrower spatial distribution, and a narrower velocity
distribution than DM particles. Differences in spatial distributions
have no effect on the dynamical mass estimates, since the estimate of
the harmonic mean radius is dominated by the influence of interlopers.
Selecting early-type galaxies helps improving the accuracy of the
virial mass estimate because less interlopers enter the sample. On the
other hand, the different velocity distribution of early-type galaxies
with respect to that of DM particles, results in $M_{\sigma}$
estimates that are too small.

\section{Discussion}\label{s-disc}
We have used a sample of 62 galaxy clusters extracted from a
cosmological simulation, each with at least 10 galaxies within the
cluster virial radius, to study the reliability of cluster mass
estimates based on the distribution of their member galaxies. Galaxies
in the simulation have been typed 'early' or 'late' based on the
average formation redshift of their stellar population. The projected
phase-space distributions of both the early- and the late-type
galaxies of the simulation have been shown to be similar to those
observed in real galaxy clusters (see Sect.~\ref{s-sims}).

Two mass estimators have been considered in our analysis. One is the
classical virial theorem estimate (corrected for the surface pressure
term), and the other is an estimate based entirely on the cluster
velocity dispersion ($M_{\sigma}$, see Sect.~\ref{s-mass}). By
application to the set of simulated clusters in full phase-space, we
have shown these estimators to be unbiased, and the clusters to be
virialized (on average) as expected for $z \sim 0$ clusters in a
$\Lambda$CDM cosmology.

In order to study how efficient are these mass estimators when applied
to observed clusters, we have analysed our set of 62 simulated
clusters in projection. Three orthogonal projections have been
considered for each cluster. DM particles or galaxies have been
selected in cylinders of given aperture radius and $\simeq 200 \,
h^{-1}$ Mpc depth, in order to simulate the effect of
interlopers. Cluster members have then been selected with methods
commonly used in the recent literature (see Sect.~\ref{s-mass}).  DM
particles have been considered as tracers of the potential, instead of
galaxies, when we wanted to explore the effects of particular
observational conditions with a sufficiently good statistics. However,
the results of the dynamical analysis are very similar when using the
same number of DM particles or galaxies, except when galaxies of a
given type are specifically selected.  Hence we think that our
conclusions based on the analyses of clusters of DM particles should
be applicable to the real world, when no distinction is made among
galaxies of different types.

Projection effects significantly affect the reliability of cluster
mass estimates, through the inclusion of interlopers among the samples
of presumed cluster members (see Sect.~\ref{s-proj}). They are
responsible for large errors in the mass estimates of some clusters,
depending on the projection direction.  However, despite few
catastrophic cases, the virial mass estimates are, on average, within
10\% of the true values, if a sufficient number of DM particles are
used as tracers of the potential, $\ga 60$. Nowadays, having a
spectroscopic sample of $\ga 60$ member galaxies per cluster does not
represent a challenging observational target, even for distant
clusters (Demarco et al. \cite{dema05}), thanks to the multiplexing
capabilities of instruments like VIMOS at the VLT (e.g. Czoske et
al. \cite{czos02}) and LRIS at Keck (e.g. Goto et al. \cite{gotal05}).
This was not the case for large spectroscopic surveys of galaxy
clusters of the past (Katgert et al. \cite{katg96}; Carlberg et
al. \cite{carl97b}).

Subclustering and the presence of groups along the line-of-sight to a
cluster are responsible for the few catastrophic cases of strong mass
under- and overestimate. Using the simple DS-test for subclustering it
is however possible to identify the worst cases and eventually remove
them from the sample (see Sect.~\ref{s-sub}). Similar conclusions have
already been reached by vH97 on their simulated clusters, and by Bird
(\cite{bird95}) observationally. Note however that it becomes increasingly
difficult to identify clusters whose masses are overestimated because
of subclustering, when the size of the data-sample is decreased. Even
with $\sim 100$ tracers of the gravitational potential per cluster,
3/4 of the cases of significant subclustering previously found using
$\sim 500$ tracers, can no longer be identified. 

The (positive) bias of virial mass estimates strongly increases as the
number of tracers of the potential is decreased below $\sim 60$ (see
Sect.~\ref{s-under}).  Since cluster velocity dispersions are always
slightly underestimated (by $\sim 5$\%) the bias has to do with an
overestimate of the harmonic mean radius. This in turn has its origin
in the presence of interlopers among the particles identified as
members. As already noticed by C97, these interlopers are
characterized by a kinematic component of their velocities that
cancels the difference in the cosmological components of the particle
and cluster velocities, i.e., these DM particles are infalling into
the cluster. Since they are outside the cluster, their spatial
distribution is not as centrally concentrated as that of real cluster
member particles, hence they contribute to increase the harmonic mean
radius estimate. Diaferio et al. (\cite{diaf99}) came to essentially the same
conclusions.

In order to reduce the mass discrepancies when $\la 60$ members per
clusters are available, one could take advantage of the fact that the
$M_{\sigma}$ estimates become {\em less} biased as the size of the
sample is decreased. I.e., if the sampling of a cluster is poor, it is
better to use $M_{\sigma}$ rather than the virial mass. This is also
the case when there is considerable uncertainty in the completeness of
the selected sample of tracers (see Sect.~\ref{s-slit}). Sampling the
same number of tracers within denser cluster regions (i.e.  reducing
the aperture size of the observational set-up) does not significantly
improve the virial mass estimate, but could be observationally more
convenient.

When using galaxies instead of DM particles as tracers, it is possible
to draw subsamples selected on the basis of the galaxy
properties. When only early-type galaxies are selected, the bias in
the virial mass estimate is strongly suppressed, even if the average
number of cluster members is very small.  A similar conclusion was
reached observationally by Biviano et al. (\cite{bivi97}), who
suggested to exclude emission-line galaxies from the sample of objects
to be used in a cluster virial mass determination (see also Sanchis et
al. \cite{sanc04}). The improvement in the virial mass estimate
obtained using only early-type galaxies depends on the fact that the
fraction of early-type galaxies is higher among cluster members than
in the field, the so-called morphology-density relation (Dressler
\cite{dres80}) which is evident also in our simulated clusters.
Hence, the interloper contamination is substantially reduced, compared
to the case in which all galaxies (or DM particles) are selected.

A comparison of our results with previous works is not
straightforward, because of several differences in the analyses. Some
of the previous works did not use the virial mass estimates at the
estimated virial radius, but only the $\sigma_v$ estimate (Frenk et
al. \cite{fren90}; vH97), or isothermal mass estimates derived from
$\sigma_v$ (C97; RB99), without using the spatial distribution of
cluster members (i.e. the harmonic mean radius estimate).  The
$M_{\sigma}$ estimate used here differs from the isothermal mass
estimates used in C97 and RB99 because their masses are derived within
fixed linear radii, and hence have a $\sigma_v^2$ dependence, while
$M_{\sigma}$ is the mass within $r_v$, and hence has a $\sigma_v^3$
dependence. Both Sanchis et al. (\cite{sanc04}) and \L okas et
al. (\cite{loka05}) derived the cluster masses from the application of
the isotropic Jeans equation. Since this method requires a sufficient
number of tracers, Sanchis et al.  (\cite{sanc04}) and \L okas et
al. (\cite{loka05}) only considered the case where 400, respectively
300, DM particles are available in each simulated cluster.  

Most previous analyses were based on the classical 3-$\sigma$ clipping
method of Yahil \& Vidal (\cite{yahi77}), while the algorithm for the
identification of interlopers used in this work is more sophisticated.
However, the methods used by Sanchis et al. (\cite{sanc04}) and \L
okas et al. (\cite{loka05}) have several similarities with the one
used in this paper. Moreover, vH97 did consider the method of den
Hartog \& Katgert (\cite{denh96}), which is part of the technique used
in this paper (see Sect.~\ref{s-mass}); vH97 indeed concluded that
this method performs better than Yahil \& Vidal's, particularly in the
presence substructures. We do not expect to see a radical change in
the average performance of the virial mass estimator as a consequence
of using a different interloper rejection technique. First of all, it
has been shown by several authors that different interloper rejection
techniques generally lead to similar cluster mass estimates, when the
number of sampled galaxies is sufficiently large (e.g. Girardi et
al. \cite{gira93}; Adami et al. \cite{adam98c}; Biviano \& Girardi
\cite{bivi03}; Pimbblet et al. \cite{pimb06}). Since a cluster
velocity dispersion decreases with radius (see Fig.~\ref{f-rvall}; see
also den Hartog \& Katgert \cite{denh96}; Biviano \& Katgert
\cite{bivi04}) an interloper rejection method based on galaxy
velocities only (like, e.g., Yahil \& Vidal's) will probably reject
more cluster members near the cluster centre, while accepting more
interlopers at the cluster edge, relative to a method that takes into
account galaxy velocities {\em and} positions. However, the radial
gradient of a cluster velocity dispersion is not strong, and the
effect of neglecting the radial variation of a cluster velocity
dispersion on the estimate of global cluster quantities, like its mass
and velocity dispersion, is likely to be negligible, {\em
unless} there is substantial substructure.  Substructure is definitely
identified much better by using the combined spatial and velocity
information than with velocity information alone (see, e.g., Girardi
\& Biviano \cite{gira02} and references therein). A detailed
comparison of various methods of interloper rejection is however
deferred to a forthcoming paper (Girardi et al., in preparation).

Both Sanchis et al. (\cite{sanc04}) and \L okas et al. (\cite{loka05})
concluded that the virial masses of clusters can be reliably estimated
using several hundreds of tracers of the gravitational potential, and
we substantially agree on this.  The method they used for estimating
cluster masses are however different from the methods we considered in
this paper. Both Sanchis et al. (\cite{sanc04}) and \L okas et
al.(\cite{loka05}, see their Figure~2) found that their method yields
a mild systematic underestimate of the cluster masses. On the other
hand, for similar numbers of tracers, our methods provide mass
estimates within $\simeq \pm 10$\% of the true values. Their method is
however aimed at deriving not only cluster masses, but also internal
velocity anisotropies of the tracers population.

The r\^ole of interlopers in the overestimate of the harmonic mean
radius was emphasized by C97 and Diaferio et al. (\cite{diaf99}), and
is in agreement with our finding. The systematic underestimate of
$\sigma_v$ by den Hartog \& Katgert's (\cite{denh96}) method was
already noted by vH97, but the effect was claimed to be larger than we
see here, $\sim 100$ km~s$^{-1}$ vs. our estimate of only $\la 50$
km~s$^{-1}$ on average. C97 and vH97 suggested that the mass estimate
bias decreases with the mass of the system, and increases with the
aperture used to select the tracers. In our analysis we find that
these trends are very marginal, if they are present at all. Perhaps
this is due to our use of a more efficient interloper rejection
algorithm.

The issue of a velocity bias of galaxies relatively to DM has been
raised several times, with some authors claiming it to be rather
strong (Frenk et al. \cite{fren96}), and others rather mild (Berlind
et al. \cite{berl03}; Faltenbacher et al. \cite{falt05}). We find that
the result depends on which galaxies are selected, since the bias is
negligible when all galaxies are considered, but significant, although
small (0.94, on average), when only early-type galaxies are considered
(see Sect.~\ref{s-gal}). The dependence of the velocity bias with
galaxy mass will be explored in a future paper.

\section{Conclusions}\label{s-summ}
Using a set of 62 clusters extracted from a $\Lambda$CDM cosmological
simulation, we investigated the behaviour of mass estimators based on
the projected phase-space of galaxies. Several observational effects
were considered, like the presence of interlopers and subclusters, the
sample size, the size of the observational aperture, incompleteness,
and the selection of different tracers of the gravitational potential
(early-type galaxies, all galaxies, DM particles).

Our results show that the virial mass estimator is almost unbiased
for samples of $\ga 60$ tracers. The average bias is 1.10,
with a scatter of 0.30 (0.40) for samples of 400 (60) cluster members.
For smaller data-sets, the bias of the virial mass estimator increases
with decreasing sample sizes, reaching a maximum of $\simeq 1.5$--1.6
when the number of cluster members is decreased to $\simeq
15$--20. The virial mass estimates can be improved by removing
clusters with significant evidence for subclustering, or by selecting
early-type galaxies as tracers. The $M_{\sigma}$ estimator, based
entirely on the cluster velocity dispersion, has a bias of $\simeq
0.90$ for sample sizes $\ga 60$ members, and is essentially unbiased
for smaller data-sets.  The scatter of the $M_{\sigma}$ estimator
increases from $\sim 0.30$ for samples of $\ga 60$ cluster members, to
$\sim 0.60$ for samples of only 10 cluster members. Since early-type
galaxies have on average a lower velocity dispersion than DM particles,
the $M_{\sigma}$ estimator based on early-type galaxies only, is biased
low.

Our analysis therefore suggests that the distribution of galaxies in
projected phase-space can be used to provide reliable estimates of a
cluster mass. In order to optimize the mass estimate, {\em early-type}
galaxies should be preferentially used as tracers, if a virial mass
estimate is required. A simpler, more robust, and less biased
estimate, especially for small data-sets, can however be obtained from
an estimate of the velocity dispersion of {\em all} galaxies
identified as cluster members. However, it is not the scope of this
paper to set up an optimized observational strategy for the
determination of cluster masses. The interloper rejection technique
adopted in our analysis has been proven to work rather efficiently,
but perhaps there is room for improvement.  We are going to compare
different interloper rejection techniques in a forthcoming
paper (Girardi et al., in preparation). Further progress is to be
expected by increasing the resolution of the simulations to check the
stability of the phase-space sampling of galaxies, and by a more
realistic galaxy classification, e.g. based on colours.

\begin{acknowledgements}
We wish to dedicate this paper to the memory of Daniel Gerbal.
We thank Gary Mamon, Lauro Moscardini and Bepi Tormen for useful
discussion and the referee for her/his useful comments.  The
simulations were carried out on the IBM-SP4 machine at the ``Centro
Interuniversitario del Nord-Est per il Calcolo Elettronico'' (CINECA,
Bologna), with CPU time assigned under a INAF/CINECA grant. This work
has been partially supported by the PD-51 INFN grant. This research
has made use of NASA's Astrophysics Data System Bibliographic
Services.
\end{acknowledgements}


\vspace*{1.0cm}

\begin{thebibliography}{}

\bibitem[2005]{abaz05} Abazajian, K., Adelman-McCarthy, J.K., Ag\"ueros, M.A. et al. 2005, \aj, 129, 1755
\bibitem[1958]{abel58} Abell, G.O. 1958, \apjs, 3, 211
\bibitem[1999]{acev99} Aceves, H., \& Perea, J. 1999, \aap, 345, 439
\bibitem[1998a]{adam98a} Adami, C., Biviano, A., \&  Mazure, A. 1998, \aap, 331, 439
\bibitem[1998c]{adam98c} Adami, C., Mazure, A., Biviano, A., Katgert, P. \&  
Rhee, G. 1998, \aap, 331, 493
\bibitem[1998b]{adam98b} Adami, C., Mazure, A., Katgert, P. \& Biviano, A. 1998, \aap, 336, 63
\bibitem[2002]{athr02} Athreya, R.M., Mellier, Y., van Waerbeke, L., et al. 2002, \aap, 384, 743
\bibitem[2002]{barr02} Barrena, R., Biviano, A., Ramella, M., Falco, E.E., \& Seitz, S. 2002, \aap, 386, 816
\bibitem[1996]{bart96} Bartelmann, M. \& Steinmetz, M. 1996, \mnras, 283, 431
\bibitem[1990]{beer90} Beers, T.C., Flynn, K., \& Gebhardt 1990, \aj, 100, 32
\bibitem[2003]{berl03} Berlind, A.A., Weinberg, D.H., Benson, A.J., et al. 2003, \apj, 593, 1 
\bibitem[1995]{bird95} Bird, C. 1995, \apj, 445, L81
\bibitem[2000]{bivi00} Biviano, A. 2000, in Constructing the Universe with Clusters of Galaxies, eds. F. Durret, \& D. Gerbal, CD-Rom, and http://nedwww.ipac.caltech.edu/level5/Biviano2/frames.html 
\bibitem[2003]{bivi03} Biviano, A., \& Girardi, M. 2003, \apj, 585, 205
\bibitem[2004]{bivi04} Biviano, A., \& Katgert, P. 2004, \aap, 424, 779
\bibitem[1992]{bivi92} Biviano, A., Girardi, M., Giuricin, G., Mardirossian, F., \& Mezzetti, M. 1992, \apj, 396, 35
\bibitem[1993]{bivi93} Biviano, A., Girardi, M., Giuricin, G., Mardirossian, F., \& Mezzetti, M. 1993, \apjl, 411, L13
\bibitem[1997]{bivi97} Biviano, A., Katgert, P., Mazure, A., et al. 1997, \aap, 321, 84
\bibitem[2002]{bivi02} Biviano, A., Katgert, P., Thomas, T., \& Adami, C. 2002, \aap, 387, 8
\bibitem[1997]{borg97} Borgani, S., Gardini, A., Girardi, M., \& Gottlober, S. 1997, New A, 2, 119
\bibitem[1999]{borg99} Borgani, S., Girardi, M., Carlberg, R.G., Yee, H.K.C., \& Ellingson, E. 1999, \apj, 527, 561
\bibitem[2004]{borg04} Borgani, S., Murante, G., Springel, V., et al. 2004, \mnras, 348, 1078
\bibitem[2005]{brad05} Brada\v{c}, M., Erben, T., Schneider, P., et al. 2005, \aap, 437, 49
\bibitem[2003]{capp03} Cappi, A., Benoist, C., da Costa, L.N., \& Maurogordato, S. 2003, \aap, 408, 905
\bibitem[1997a]{carl97a} Carlberg, R.G., Yee, H.K.C., \& Ellingson, E. 1997a, \apj, 478, 462
\bibitem[1997b]{carl97b} Carlberg, R.G., Yee, H.K.C., Ellingson, E., et al. 1997b, \apj, 476, L7
\bibitem[1997]{cen97} Cen, R. 1997, \apj, 485, 39 (C97)
\bibitem[2001]{clow01} Clowe, D., Trentham, N., \& Tonry, J. 2001, \aap, 369, 16
\bibitem[2004]{clow04} Clowe, D., De Lucia, G., \& King, L. 2004, \mnras, 350, 1038
\bibitem[2000]{coli00} Col\'{\i}n, P., Klypin, A.A., \& Kravtsov, A.V. 2000, \apj, 539, 561
\bibitem[2004]{cypr04} Cypriano, E.S., Sodr\'e, L. Jr., Kneib, J.-P., \& Campusano, L.E. 2004, \apj, 613, 95
\bibitem[2002]{czos02} Czoske, O., Moore, B., Kneib, J.-P., \& Soucail, G. 2002, \aap, 386, 31
\bibitem[1992]{dalt92} Dalton, G.B., Efstathiou, G., Maddox, S.J., \& Sutherland, W.J. 1992, \apjl, 390, L1
\bibitem[1980]{dane80} Danese, L., de Zotti, G., \& di Tullio, G. 1980, \aap, 82, 322
\bibitem[1981]{dane81} Danese, L., de Zotti, G., Giuricin, G., Mardirossian, F., Mezzetti, M., \& Ramella, M. 1981, \apj, 244, 777
\bibitem[2005]{dema05} Demarco, R., Rosati, P., Lidman, C., et al. 2005, \aap, 432, 381
\bibitem[1996]{denh96} den Hartog, R., \& Katgert, P. 1996, \mnras, 279, 349
\bibitem[1999]{diaf99} Diaferio, A., Kauffmann, G., Colberg, J.M., \& White, S.D.M. 1999, \mnras, 307, 537
\bibitem[2005]{diaf05} Diaferio, A., Geller, M.J., \& Rines, K.J. 2005, \apj, 628, L97
\bibitem[2004]{diem04} Diemand, J., Moore, B., \& Stadel, J. 2004, \mnras, 353, 624
\bibitem[2004]{dola04} Dolag, K., Bartelmann, M., Perrotta, F., et al. 2004, \aap, 416, 853
\bibitem[1980]{dres80} Dressler, A. 1980, \apj, 236, 351
\bibitem[1988]{dres88} Dressler, A., \& Shectman, S.A. 1988, \aj, 95, 985 (DS)
\bibitem[2001]{dye01} Dye, S., Taylor, A.N., Thommes, E.M., et al. 2001, \mnras, 321, 685
\bibitem[2005]{falt05} Faltenbacher, A., Allgood, B., Gottl\"ober, S., Yepes, G.,\& Hoffman, Y. 2005, \mnras, 362, 1099
\bibitem[1990]{fren90} Frenk, C., White, S.D.M., Efstathiou, G., \& Davis, M. 1990, \apj, 351, 10
\bibitem[1996]{fren96} Frenk, C., Evrard, A.E., White, S.D.M., \& Summers, F.J. 1996, \apj, 472, 460
\bibitem[2004]{gao04} Gao, L., De Lucia, G., White, S.D.M., \& Jenkins, A. 2004, \mnras, 352, L1
\bibitem[2005]{gava05} Gavazzi, R. 2005, \aap, 443, 793
\bibitem[1998]{ghig98} Ghigna, S., Moore, B., Governato, F., et al. 1998, \mnras, 300, 146
\bibitem[2000]{ghig00} Ghigna, S., Moore, B., Governato, F., et al. 2000, \apj, 544, 616
\bibitem[2002]{gira02} Girardi, M. \& Biviano, A. 2002, in Merging Processes in Galaxy Clusters, eds. L. Feretti, I.M. Gioia, \& G. Giovannini (Dordrecht: Kluwer), 39
\bibitem[1993]{gira93} Girardi, M., Biviano, A., Giuricin, G., Mardirossian, F., \& Mezzetti, M. 1993, \apj, 404, 38
\bibitem[1996]{gira96} Girardi, M., Fadda, D., Giuricin, G., et al. 1996, \apj, 457, 61
\bibitem[1998]{gira98} Girardi, M., Giuricin, G., Mardirossian, F., Mezzetti, M., \& Boschin, W. 1998, \apj, 505, 74
\bibitem[2005]{glad05} Gladders, M.D., \& Yee, H.K.C. 2005, \apjs, 157, 1
\bibitem[2005]{goto05} Goto, T. 2005, \mnras, 359, 1415
\bibitem[2005]{gotal05} Goto, T., Postman, M., Cross, N.J.G., et al. 2005, \apj, 621, 188 2005
\bibitem[1996]{haar96} Haardt, F., \& Madau, P. 1996, \apj, 461, 20
\bibitem[1996]{katg96} Katgert, P., Mazure, A., Perea, J. et al. 1996, \aap, 310, 8
\bibitem[1998]{katg98} Katgert, P., Mazure, A., den Hartog, R., et al. 1998, \aaps, 129, 399
\bibitem[2004]{katg04} Katgert, P., Biviano, A. \& Mazure, A. 2004, \apj, 600, 657
\bibitem[2004]{kay04} Kay, S.T., Thomas, P.A., Jenkins, A., \& Pearce, F. 2004, \mnras, 355, 1091
\bibitem[1999]{klyp99} Klypin, A., Gottl\"ober, S., Kravtsov, A.V., \& Khokhlov, A.M. 1999, \apj, 516, 530
\bibitem[1998]{kora98} Koranyi, D.M., Geller, M.J., Mohr, J.J., \& Wegner, G. 1998, AJ, 116, 2108
\bibitem[1960]{limb60} Limber, D.N., \& Mathews, W.G. 1960, \apj, 132,
  286
\bibitem[2003]{lin03} Lin, J.-T., Mohr, J.J.,, \& Stanford, S.A 2003 \apj, 591, 749
\bibitem[2003]{loka03} \L okas, E.L. \& Mamon, G.A. 2003, \mnras, 343, 401
\bibitem[2005]{loka05} \L okas, E.L. Wojtak, R., Gottl\"ober, S., Mamon, G.A. \& Prada, F. 2005, astro-ph/0511723
\bibitem[2005]{lomb05}  Lombardi, M., Rosati, P., Blakeslee, J.P., et al. 2005, \apj, 623, 42
\bibitem[2004]{lope04} Lopes, P.A.A., de Carvalho, R.R., Gal, R.R., et al. 2004, \aj, 128, 1017
\bibitem[1992]{lums92} Lumsden, S.L., Nichol, R.C., Collins, C.A., \& Guzzo, L. 1992, \mnras, 258, 1
\bibitem[1983]{luce83} Lucey, J.R. 1983, \mnras, 204, 33
\bibitem[2005]{merc05} Merch\'an, M.E., \& Zandivarez, A. 2005, \apj, 630, 759
\bibitem[2001]{metz01} Metzler, C.A., White, M., \& Loken, C. 2001, \apj, 547, 560
\bibitem[1999]{metz99} Metzler, C.A., White, M., Norman, M., \& Loken, C. 1999, \apj, 520, L9
\bibitem[2005]{mill05} Miller, C.J., Nichol, R.C., Reichart, D., et al. 2005, \aj, 130, 968 
\bibitem[2003]{macc03} Macci\`o, A.V., Murante, G., \& Bonometto, S.P. 2003, \apj, 588, 35
\bibitem[1977]{moss77} Moss, C. \& Dickens, R.J. 1977, \mnras, 178, 701
\bibitem[2004]{mura04} Murante, G., Arnaboldi, M., Gerhard, O., et al. 2004, \apj, 607, L83
\bibitem[2005]{naga05} Nagai, D., \& Kravtsov, A.V. 2005, \apj, 618, 557
\bibitem[1997]{nava97} Navarro, J.F., Frenk, C.S., \& White, S.D.M. 1997, \apj, 490, 493 (NFW)
\bibitem[2005]{ogu05} Oguri, M., Takada, M., Umetsu, K., \& Broadhurst, T. 2005, \apj, 632, 841
\bibitem[2004]{ota04} Ota, N., Pointecouteau, E., Hattori, M., \& Mitsuda, K. 2004, \apj, 601, 120
\bibitem[1990]{pere90} Perea, J., del Olmo, A., \& Moles, M. 1990, \aap, 237, 319
\bibitem[2005]{pimb06} Pimbblet, K.A., Smail, I., Edge, A.C., O'Hely, E.,
Couch, W.J., \& Zabludoff, A.I. 2006, \mnras, 366, 645 
\bibitem[2005]{pope05} Popesso, P., Biviano, A., B\"ohringer, H., Romaniello, M., \& Voges, W. 2005, \aap, 433, 431
\bibitem[2006]{pope06} Popesso, P., Biviano, A., B\"ohringer, H., Romaniello, M., \& Voges, W. 2006, \aap, 445, 29
\bibitem[2001]{rame01} Ramella, M., Boschin, W., Fadda, D., \& Nonino, M. 2001, \aap, 368, 776
\bibitem[2004]{rame04} Ramella, M., Boschin, W., Geller, M.J., Mahdavi, A., \& Rines, K. 2004, \aj, 128, 2022
\bibitem[2004]{rasi04} Rasia, E., Tormen, G., \& Moscardini, L. 2004, \mnras, 351, 237
\bibitem[1999]{rebl99} Reblinsky, K., \& Bartelmann, M. 1999, \aap, 345, 1 (RB99)
\bibitem[2005]{reed05} Reed, D., Governato, F., Quinn, T., et al. 2005, \mnras, 359, 1357
\bibitem[2001]{rick01} Ricker, P.M., \& Sarazin, C.L. 2001, \apj, 561, 621
\bibitem[2004]{rine04} Rines, K., Geller, M.J., Diaferio, A., Kurtz, M.J., \& Jarrett, T.H. 2004, \aj, 128, 1078
\bibitem[2002]{rosa02} Rosati, P., Borgani, S., \& Norman, C. 2002, \araa, 40, 539
\bibitem[2004]{sanc04} Sanchis, T., \L okas, E.L., \& Mamon, G.A. 2004, \mnras, 347, 1193
\bibitem[1993]{schi93} Schindler, S. \& M\"uller, E. 1993, \aap, 272, 137
\bibitem[2002]{smit02} Smith, G.P., Smail, I., Kneib, J.-P., et al. 2002, \mnras, 333, L16
\bibitem[1989]{sodr89} Sodr\'e, L.Jr., Capelato, H.V., Steiner, J.E., \& Mazure, A. 1989, \aj, 97, 1279
\bibitem[2005]{spri05} Springel, V. 2005, \mnras, 364, 1105  
\bibitem[2003]{spri03} Springel, V., \& Hernquist, L. 2003, \mnras, 339, 289
\bibitem[2001]{spri01} Springel, V., Yoshida, N., \& White, S.D.M. 2001, New A, 6, 79  
\bibitem[2001]{stad01} Stadel, J.G. 2001, PhD Thesis Univ. Washington
\bibitem[1986]{the86} The, L.S., \& White, S.D.M.  1986, \aj, 92, 1248
\bibitem[2006]{thom06} Thomas, T., \& Katgert, P. 2006, \aap, 446, 19
\bibitem[1997]{torm97} Tormen, G., Bouchet, F.R., \& White, S.D.M. 1997, \mnras, 286, 865
\bibitem[1997]{vanh97} van Haarlem, M.P., Frenk, C.S., \& White, S.D.M. 1997, \mnras, 287, 817 (vH97)
\bibitem[1997]{vank97} van Kampen, E.., \& Katgert, P. 1997, \mnras, 289, 327 
\bibitem[2005]{voit05} Voit, G.M. 2005, AdSpR, 36, 701
\bibitem[2004]{wamb04} Wambsganss, J., Bode, P., \& Ostriker, J.P. 2004, \apj, 603, L93
\bibitem[2005]{wamb05} Wambsganss, J., Bode, P., \& Ostriker, J.P.  2005, \apj, 635, L1
\bibitem[2000]{wu00} Wu, X.-P. 2000, \mnras, 316, 299
\bibitem[1977]{yahi77} Yahil, A., \& Vidal, N.V. 1977, \apj, 214, 347
\bibitem[2005]{yang05} Yang, X., Mo, H.J., van den Bosch, F., Weinmann, S.M.,
Li, C., \& Jing, Y.P. 2005, \mnras, 362, 711
\bibitem[1933]{zwic33} Zwicky, F. 1933, Helv. Phys. Acta, 6, 110

\end{thebibliography}
\end{document}